\documentclass{aa}
\usepackage[varg]{txfonts}
\usepackage{natbib}
\usepackage{hyperref}
\usepackage{orcidlink}

\usepackage{upgreek}
\usepackage{graphicx}   
\usepackage{subcaption}

\newcommand{\src}{{SXP31.0}\xspace}
\newcommand{\nustar}{\textit{NuSTAR}\xspace}
\newcommand{\swift}{\textit{Swift}\xspace}

\newcommand{\rxte}{\textit{RXTE}\xspace}

\newcommand{\srg}{\textit{SRG}\xspace}

\newcommand{\msun}{M_{\odot}}

\def\flux{erg\,s$^{-1}$\,cm$^{-2}$\xspace}
\def\lum{erg\,s$^{-1}$\xspace}

\begin{document}

\title{Discovery of a 0.8-mHz quasi-periodic oscillation in the transient X-ray pulsar SXP31.0 and associated timing  transitions}
\titlerunning{Discovery of 0.8 mHz QPO in \src}

\authorrunning{Salganik, A., et al.}

\author{Alexander~Salganik\inst{\ref{in:UTU}}\orcidlink{0000-0003-2609-8838} 
\and    Sergey~S.~Tsygankov\inst{\ref{in:UTU}}\orcidlink{0000-0002-9679-0793} 
\and    Sergey~V.~Molkov\inst{\ref{in:IKI}} \orcidlink{0000-0002-5983-5788}
\and Igor Yu. Lapshov \inst{\ref{in:IKI}}\orcidlink{0009-0000-4769-452X}  
\and Alexander~A.~Lutovinov\inst{\ref{in:IKI}}\orcidlink{0000-0002-6255-9972}   
\and Alexey Yu. Tkachenko\inst{\ref{in:IKI}}\orcidlink{0000-0002-7486-1730}
\and  Alexander A. Mushtukov\inst{\ref{in:OXF}}\orcidlink{0000-0003-2306-419X}   
\and Juri Poutanen\inst{\ref{in:UTU}}\orcidlink{0000-0002-0983-0049}}
\institute{
Department of Physics and Astronomy, 20014 University of Turku,  Finland 
\label{in:UTU} \\ \email{alsalganik@gmail.com}
\and
Space Research Institute of the Russian Academy of Sciences, 117997 Moscow, Russia
\label{in:IKI}
\and
Astrophysics, Department of Physics, University of Oxford, Denys Wilkinson Building, Keble Road, Oxford OX1 3RH, UK
\label{in:OXF}
}

   \date{Received 24 June 2025 / Accepted 14 October 2025}

\abstract{
We present the first broadband spectral and timing study of the Be/X-ray pulsar XTE\,J0111.2$-$7317 (SXP31.0) during the first major outburst since its discovery in 1998. This giant type~II outburst, observed between April and September 2025, marks the source's return to activity after nearly three decades of quiescence. Using \nustar\ observations together with data from \swift/XRT and \srg/ART-XC, we followed the outburst’s evolution, with the source reaching a bolometric luminosity of $L_{\rm bol} = 3.6 \times 10^{38}$~\lum. The broadband spectra are well described by an absorbed cutoff power law, two blackbody components (hot and soft), and a narrow Fe~K$\alpha$ line. No cyclotron absorption features were detected in either the phase-averaged or phase-resolved spectra in the 5--50 keV band.
Most notably, we report the discovery of a previously undetected quasiperiodic oscillation (QPO) at $0.8 \pm 0.1$~mHz, characterized by a fractional root-mean-square (rms) amplitude of 14\% at a super-Eddington bolometric luminosity of $L_{\rm bol} = 2.5 \times 10^{38}$~\lum. In contrast, the previously reported 1.27~Hz QPO was not detected. While the 0.8~mHz QPO is present, the pulsed fraction (PF) is low in soft X-rays, which is consistent with other super-Eddington pulsars exhibiting mHz QPOs; however, it rises above 20~keV to reach 35\%. 
The QPO vanishes in subsequent observations coinciding with a sharp increase in the PF and a distinct change in pulse profile morphology. It was not observed in any follow-up observations at luminosities above or below its initial detection, suggesting it is a transient phenomenon.}


\maketitle

\section{Introduction}

High-mass X-ray binaries (HMXBs) are systems in which a compact object, either a neutron star (NS) or a black hole, accretes matter from a massive companion. Many HMXBs host strongly magnetized NSs (with a magnetic field strength of $B$$ \sim$$10^{12}$--$10^{13}$\,G) and, therefore, they often exhibit X-ray pulsations. Among HMXBs, Be/X-ray binaries (BeXRBs) span an exceptionally wide range of accretion rates, providing valuable laboratories for investigating accretion processes, ultrastrong magnetic fields, and quantum electrodynamics \citep[see][for a recent review]{MushtukovTsygankov2024}. 
In BeXRBs, the accreting  matter is supplied through the circumstellar decretion disk around the rapidly rotating Be companion \citep[see][for a review]{Reig2011}. As a result, these systems exhibit two types of activity: regular type~I outbursts occurring near periastron passage, with typical X-ray luminosities of $\sim10^{37}$~\lum and giant type~II outbursts that exceed this level by one or more orders of magnitude.

One such system is the transient source XTE~J0111.2$-$7317, which was discovered by the \textit{Rossi X-ray Timing Explorer} (\rxte) in November 1998 during a type~II outburst, when it reached a luminosity exceeding $10^{38}$~\lum \citep{Chakrabarty1998}.
Coherent X-ray pulsations with a period of approximately 31~s were detected, confirming its classification as an accreting X-ray pulsar  \citep[XRP;][]{Chakrabarty1998}. For convenience, we adopted the short designation \src, following the naming convention proposed by \citet{Coe2005}. \src was confirmed to be located  in the Small Magellanic Cloud \citep[SMC;][]{Coe1998, Yokogawa2000}. Simultaneous observations by the \textit{Compton Gamma-Ray Observatory} (\textit{CGRO}) detected hard X-ray emission \citep{Wilson1998} and follow-up observations with \textit{Advanced Satellite for Cosmology and Astrophysics}  (\textit{ASCA}) have revealed a pulsating soft X-ray excess. Observations during a quiescent phase in September 2009 showed the source at a luminosity of $L_{\rm X} \approx 1.8 \times 10^{34}$~\lum \citep{Christodoulou2016}.

The optical counterpart to \src\ was identified as 2MASS J01110860$-$7316462 \citep{Coe2000} and the companion was classified as a B0.5--1Ve star, confirming the system as a BeXRB \citep{Covino2001, Coe2003}. The orbital period  of \src was determined to be $90.5 \pm 0.1$~d \citep{Rajoelimanana2011, Bird2012}. Early H$\alpha$ imaging revealed a compact, irregular emission nebula surrounding the star, initially interpreted as a possible supernova remnant \citep{Coe2000},  although spectroscopy would later identify it more plausibly as a locally photoionized \ion{H}{ii} region \citep{Coe2003}.

To date, observations of \src have not yielded a direct measurement of its magnetic field strength, as cyclotron resonant scattering features (CRSFs) have not been detected in its X-ray spectrum \citep[see, e.g.,][]{Staubert2019}. Using \rxte data from the 1998 outburst, \citet{Kaur2007} discovered a quasi-periodic oscillation (QPO) at 1.27~Hz and estimated the NS magnetic field strength to be (2--4)$\times 10^{12}$~G based on the magnetospheric beat frequency model (BFM). The field strength of \src\ thus remains uncertain and can only be inferred through indirect methods.

Recent observations with the \textit{Neil Gehrels Swift Observatory}  (\swift) indicate renewed activity from \src. The \swift SMC Survey (S-CUBED) detected enhanced X-ray emission from the previously quiescent source on 2025 April 15, with a luminosity of $\approx1.8 \times 10^{37}$~\lum in the 0.3--10~keV band \citep{Gaudin2025}, assuming a distance of 62.44~kpc to the SMC \citep{Graczyk2020}. The source was not detected in the prior survey epoch on March 11, suggesting the onset of a new transient outburst. Subsequently, on 2025 April 26  the Mikhail Pavlinsky ART-XC telescope on board the \textit{Spectrum Roentgen Gamma} (\srg) observatory detected a bright transient at a position consistent with \src, with an X-ray flux of $(1.2 \pm 0.2) \times 10^{-10}$ \flux in the 4--12 keV band \citep{Semena2025}, which corresponds to the luminosity of $\sim5 \times 10^{37}$~\lum, consistent with the onset of a Type~II outburst.

The Be/X-ray pulsar \src has previously exhibited outbursts with luminosities of $\sim 10^{38}$~\lum, approaching the Eddington limit, $L_{\rm Edd}$ for an NS, making it a promising candidate for studying accretion at high rates. Only a handful of super-Eddington pulsars have been identified to date and their properties remain poorly understood. More broadly, XRPs accreting near or above $L_{\rm Edd}$ provide a valuable opportunity to explore the transition between standard accretion regimes and the radiation-pressure-dominated flows typical of ultraluminous X-ray sources (ULXs), as well as the role of magnetic field interactions in this regime.

In this paper, we present the first comprehensive spectral and timing study of \src in the super-Eddington accretion regime, based on observations obtained during its 2025 type~II outburst. We triggered dedicated \textit{Nuclear Spectroscopic Telescope Array} (\nustar) and \srg/ART-XC Target of Opportunity (ToO) observations, supported by regular \swift monitoring.  Among our key findings is the identification of a previously unreported 0.8~mHz QPO, detected in the early phase of the outburst. The same feature was independently detected by \citet{Roy2025}, based on the same \nustar ToO dataset. Our preliminary report can be found in \citet{2025ATel17216....1S}. Here, we present the first detailed spectral and timing analysis and trace the QPO evolution across multiple epochs and luminosity states.

\section{Observations and data reduction}
\label{sec:data}
Our analysis is based primarily on broadband \nustar and \srg/ART-XC observations of \src, which provide wide energy coverage. These data are supplemented by the \swift data to trace the soft X-ray spectral evolution. The observation log is provided in Table~\ref{table:all_obs}.

\begin{table} 
\centering
\caption{Observations used in this work.}
\label{table:all_obs}
\begin{tabular}{lcccc}
\hline\hline
ObsID & Instrument & Start &  End  & Exposure  \\
  &   & (MJD) &   (MJD) &  (ks) \\
\hline
00019718001 & \swift/XRT& 60781.26 & 60781.47 & 5.48 \\
00019718002 & \swift/XRT& 60788.86 & 60788.93 & 1.84 \\
00019718003 & \swift/XRT& 60789.71 & 60789.91 & 4.92 \\
00019718004 & \swift/XRT& 60792.11 & 60792.50 & 3.58 \\
00019718005 & \swift/XRT& 60792.56 & 60792.70 & 0.74 \\
ArtObs1     & \srg/ART-XC& 60795.94 & 60796.71 & 66.49\\
91101311002      & \nustar & 60797.82 & 60798.86 & 80.08\\
03400041001 & \swift/XRT& 60798.03 & 60798.04 & 1.05 \\
00019718007 & \swift/XRT& 60798.30 & 60798.30 & 0.50 \\
00019718008 & \swift/XRT& 60799.07 & 60799.86 & 4.27 \\
00019718009 & \swift/XRT& 60802.05 & 60802.97 & 1.98 \\
00019718010 & \swift/XRT& 60804.01 & 60805.12 & 2.88 \\
ArtObs2     & \srg/ART-XC& 60812.87 & 60813.87 & 85.57\\
00019718011 & \swift/XRT& 60818.39 & 60818.98 & 2.69 \\
00019718012 & \swift/XRT& 60819.23 & 60819.96 & 4.58 \\
00019718013 & \swift/XRT& 60822.22 & 60822.95 & 4.63 \\
00019718014 & \swift/XRT& 60825.33 & 60826.00 & 4.48 \\
91101314002     & \nustar & 60825.98 & 60826.49 & 55.20 \\
00019718015 & \swift/XRT& 60828.06 & 60828.98 & 3.93 \\
00019718016 & \swift/XRT& 60831.04 & 60831.90 & 3.22 \\
ArtObs3     & \srg/ART-XC& 60832.45 & 60833.53 & 93.85 \\
00019718017 & \swift/XRT& 60834.03 & 60834.75 & 4.58 \\
00019718018 & \swift/XRT& 60837.08 & 60837.93 & 4.80 \\
00019718019 & \swift/XRT& 60840.06 & 60840.80 & 4.42 \\
00019718020 & \swift/XRT& 60844.54 & 60845.86 & 5.13 \\
ArtObs4     & \srg/ART-XC& 60845.99 & 60847.92 & 167.34 \\
00019718021 & \swift/XRT& 60846.04 & 60846.70 & 5.23 \\
00019718022 & \swift/XRT& 60849.41 & 60850.00 & 4.84 \\
00019718023 & \swift/XRT& 60852.39 & 60852.72 & 2.46 \\
00019718024 & \swift/XRT& 60853.69 & 60853.83 & 1.79 \\
00019718025 & \swift/XRT& 60868.71 & 60868.78 & 1.76 \\
00019718026 & \swift/XRT& 60870.78 & 60870.86 & 2.53 \\
00019718027 & \swift/XRT& 60873.18 & 60873.84 & 4.41 \\
00019718028 & \swift/XRT& 60876.03 & 60876.76 & 4.47 \\
00019718032 & \swift/XRT& 60888.41 & 60888.42 & 0.77 \\
00019718033 & \swift/XRT& 60891.40 & 60891.41 & 0.39 \\
00019718036 & \swift/XRT& 60900.34 & 60900.35 & 0.95 \\
00019718038 & \swift/XRT& 60915.99 & 60916.00 & 0.50 \\
00019718039 & \swift/XRT& 60918.58 & 60918.58 & 0.68 \\
00019718041 & \swift/XRT& 60924.15 & 60924.16 & 0.52 \\
00019718043 & \swift/XRT& 60930.50 & 60930.50 & 0.61 \\
\hline
\end{tabular}
\end{table}

\subsection{\nustar observatory}

\nustar comprises two identical, co-aligned X-ray focal plane modules, FPMA and FPMB \citep{Harrison2013}, which provide sensitive coverage in the 3--79~keV energy band. The \nustar observations of \src\ presented here were obtained through a ToO request submitted by our team following the initial identification of the ongoing outburst. Observations were performed on May 2--3, 2025  (ObsID 91101311002; MJD 60797--60798) and May 30--31, 2025  (ObsID 91101314002; MJD 60825--60827), hereafter referred to as NuObs1 and NuObs2, respectively. The source events were extracted using a circular region with a 50\arcsec\ radius centered on the position reported by \citet{Gaudin2025}: RA (J2000) = 01$^{\mathrm{h}}$11$^{\mathrm{m}}$$09\fs51$ (17\fdg78962), Dec (J2000) = $-73\degr16\arcmin44\farcs9$ ($-73\fdg27915$), with a 90\% confidence error radius of 6.3\arcsec. A 100\arcsec\ radius background region was chosen to optimize the signal-to-noise ratio, particularly at higher energies. The background was extracted from a source-free region in the same detector quadrant (near a detector corner).

\begin{table*} 
\centering
\caption{Observations grouped by analysis epoch.}
\label{table:periods}
\begin{tabular}{llccc}
\hline\hline
 & Observation & Observation Start (MJD) & Observation End (MJD) & Period (s) \\
\hline
Epoch 1 & ArtObs1     & 60795.94 & 60796.71 & $30.4520(1)$\\
& NuObs1      & 60797.82 & 60798.86 & 30.44970(4)\\
& \multicolumn{4}{l}{+ \swift/XRT (00019718007, 03400041001)} \\
\hline
Epoch 2 & ArtObs2     & 60812.87 & 60813.87 & 30.42116(2)\\
& \multicolumn{4}{l}{+ \swift/XRT (00019718011)} \\
\hline
Epoch 3 & NuObs2      & 60825.98 & 60826.49 & 30.39372(3)\\
& \multicolumn{4}{l}{+ \swift/XRT (00019718014)} \\
\hline
Epoch 4 & ArtObs3      & 60832.45 & 60833.53 & 30.38156(3)\\
& \multicolumn{4}{l}{+ \swift/XRT (00019718016, 00019718017)} \\
\hline
Epoch 5 & ArtObs4      & 60845.99 & 60847.92  & 30.35945(2)\\
& \multicolumn{4}{l}{+ \swift/XRT (00019718021)} \\
\hline
\end{tabular}
\end{table*}

Data reduction of the \nustar observations was carried out following the standard processing guidelines.\footnote{\url{https://heasarc.gsfc.nasa.gov/docs/nustar/analysis/nustar_swguide.pdf}} The data were processed using \textsc{HEASoft} v6.35.1 and CALDB version 20250415 (including clock correction file 20100101v204). Event files from FPMA and FPMB were barycenter-corrected using the \texttt{nuproducts} tool (via the built-in \texttt{barycorr} option). No significant ghost-ray and stray-light contamination from the nearby bright source SMC~X-1 was detected in either FPMA or FPMB for both NuObs1 and NuObs2. Solar activity was elevated during NuObs2, so we filtered outlying events following official recommendations.\footnote{\url{https://github.com/NuSTAR/nustar-gen-utils/blob/main/notebooks/GTI\_filter\_solar\_flare.ipynb}} The spectra and the light curves were extracted with the \texttt{nuproducts} tool as part of the \texttt{nustardas} pipeline. Background-subtracted light curves from FPMA and FPMB were combined using \texttt{lcmath} to improve statistics. No binary motion correction was applied, as the complete set of the source orbital parameters  remains unknown.

\begin{figure}
\centering
\includegraphics[width=0.95\columnwidth]{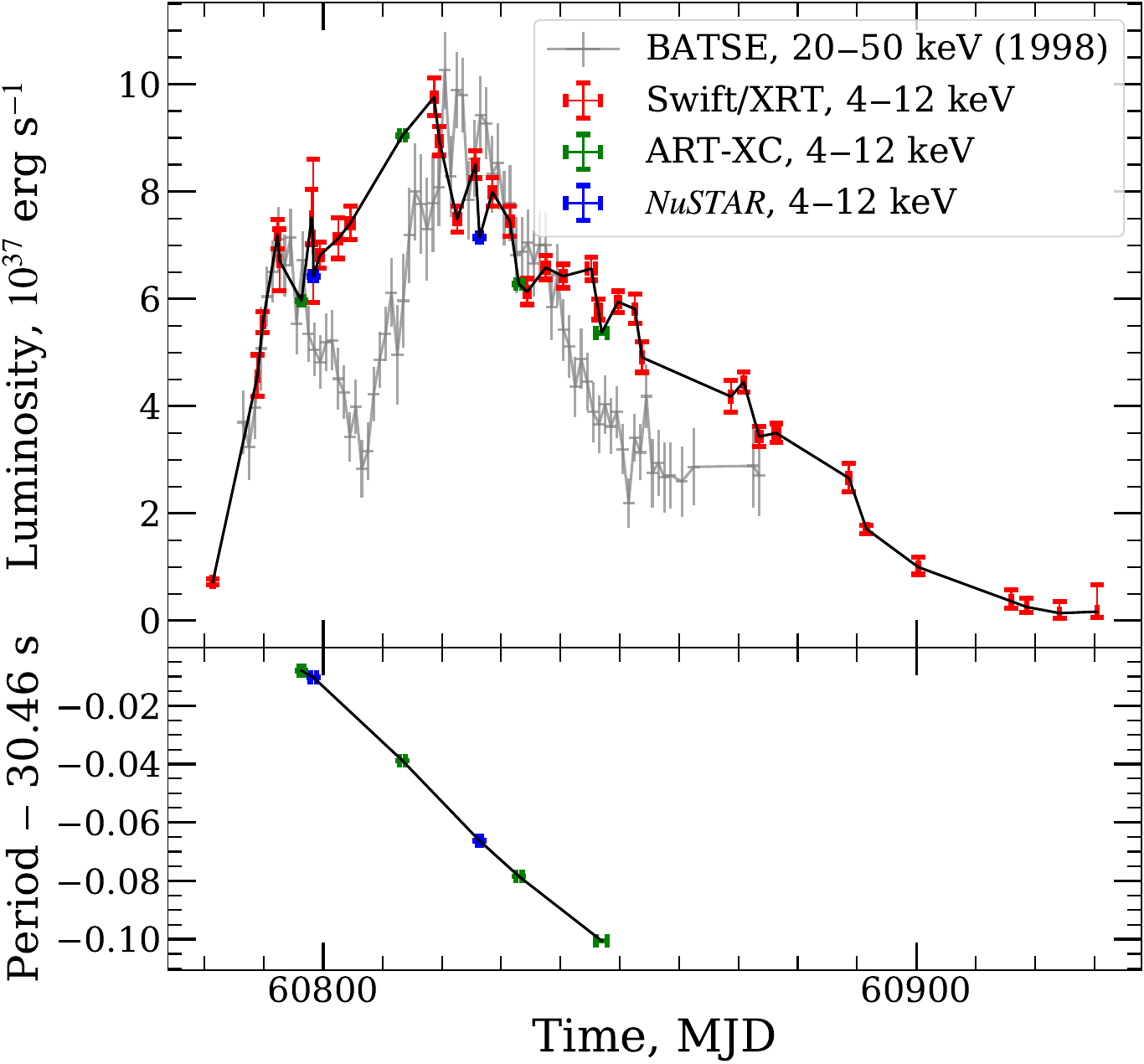}
\caption{Evolution of \src during the 2025 outburst based on \swift/XRT, \nustar, and \srg/ART-XC observations. 
The light curve in the 4--12~keV energy range is shown at the top panel. 
The gray points show the \textit{CGRO}/BATSE 20--50~keV pulsed luminosity from the 1998 outburst, overplotted for comparison. Along the time axis it was placed arbitrarily, and the luminosity was scaled by a factor of 0.45. 
The spin period (see Table~\ref{table:periods}) is shown at the bottom panel.}
\label{fig:lightcurve}
\end{figure}

\subsection{\textit{SRG}/ART-XC}
The Mikhail Pavlinsky ART-XC telescope is one of two telescopes installed aboard the \srg observatory, which was launched in 2019 from the Baikonur cosmodrome to the L2 libration point of the Sun-Earth system \citep{2021A&A...656A.132S}. 
The {ART-XC} telescope is an imaging instrument consisting of seven modules and operating in photon-counting mode \citep{2021A&A...650A..42P} in the 4--30 keV energy band, but the detectors can operate up to $\sim 120$~keV and the highest sensitivity is achieved at 4--12 keV.
The time resolution of the instrument is $\sim23$\,$\mu$s and the angular resolution is $\sim53$\arcsec.
The data were processed using the \textsc{artproducts} v1.0 software with the latest CALDB {v20230228}.
For the spectral and timing analysis, we extracted photons from a circle with a radius of 1\farcm8 around the source position and we also applied an energy filtering when needed. Observations  performed on 2025 April 30--May 1  (MJD 60795--60796), May 17--18 (MJD 60812--60813),  June 6--7 (MJD 60832--60833), and June 19--21 (MJD 60845--60847)  are referred to below as ArtObs1, ArtObs2, ArtObs3, and ArtObs4, respectively. 
The background was estimated by collecting photons from a circular region with a radius of 5\farcm4 on the detectors that did not overlap with the region in which the source photons were extracted. \src was the only detected source within the ART-XC field of view during the observation.

\subsection{\swift observatory}

To track the evolution of the outburst (see Fig.~\ref{fig:lightcurve}), we used data from the monitoring campaign with the XRT telescope on board the \swift observatory \citep{Burrows2005,Gehrels2004}, comprising 35 observations (see Table~\ref{table:all_obs}). These observations span MJD 60781.5--60930.5 and were carried out in photon counting (PC) or windowed timing (WT) modes, chosen according to the source count rate. The source and background spectra for each individual observation were extracted using the UK Swift Science Data Centre’s XRT data-analysis software,\footnote{\url{https://www.swift.ac.uk/user_objects/}} which also automatically selected the extraction regions following the method described in \citet{Evans2009}.

\subsection{Spectral data approximation}
\label{sec:spectral_data}

To enable a joint spectral fitting, the broadband spectra were grouped into five distinct epochs based primarily on similar flux levels. Observations within each group were taken at comparable luminosities and, coincidentally, they also exhibit consistent pulse profile morphology (see Sects.~\ref{sec:timing} and \ref{sec:timing_noqpo}). Within each epoch, data from \nustar, \srg/ART-XC, and \swift/XRT were combined. All spectra were binned with a minimum of one count per energy bin and fitted using W-statistics \citep{Wachter1979}.

Unless stated otherwise, all uncertainties correspond to the $1\sigma$ confidence level, and all fluxes reported in this work are unabsorbed. Luminosities were computed assuming a distance of 62.44~kpc to the SMC \citep{Graczyk2020}.

To construct the long-term light curve, individual \swift/XRT spectra were fitted with an absorbed power law plus blackbody model, $\texttt{tbabs} \times (\texttt{po} + \texttt{bbodyrad})$, with the blackbody temperature fixed at $T=0.22$~keV (see Sect.~\ref{sec:spectrum}). Throughout this analysis, the hydrogen column density was fixed at $N_{\mathrm{H}} = 0.18 \times 10^{22}~\mathrm{cm^{-2}}$, following \citet{Yokogawa2000}. The resulting long-term light curve is shown in Fig.~\ref{fig:lightcurve}. For comparison, the \textit{CGRO}/BATSE 20--50~keV pulsed flux from the 1998 outburst\footnote{\url{https://gammaray.nsstc.nasa.gov/batse/pulsar/data/sources/xtej0111p2.html}} \citep{Yokogawa2000} is also overplotted. For the conversion from Crab units to the energy flux in 20--50~keV range, we adopted $1~\mathrm{Crab}=9.22\times10^{-9}$ \flux \citep{Frontera2007}. The BATSE light curve of the 1998 outburst differs in shape from the 2025 event: while both reached comparable peak luminosities of $\sim10^{38}$~\lum, the 1998 outburst exhibited a pronounced two-humped shape and its decay was noticeably steeper. However, these comparisons should be taken with caution since the BATSE measurements represent pulsed flux.

\section{Results}

We performed a detailed timing analysis, including energy-resolved pulse profiles and pulsed fraction measurements, and we report the discovery and modeling of a transient 0.8~mHz QPO. We also present results of the broadband spectral fitting. To investigate the evolution throughout the outburst, the observations were divided into two main intervals: a QPO-active phase (epoch~1; see Table~\ref{table:periods}), when the 0.8~mHz QPO was detected, and a subsequent QPO-free phase (epochs~2--5).

\subsection{Timing analysis during QPO-active interval}
\subsubsection{Detection of the 0.8 mHz QPO}

\begin{figure} 
\centering
\includegraphics[width=1.00\linewidth]{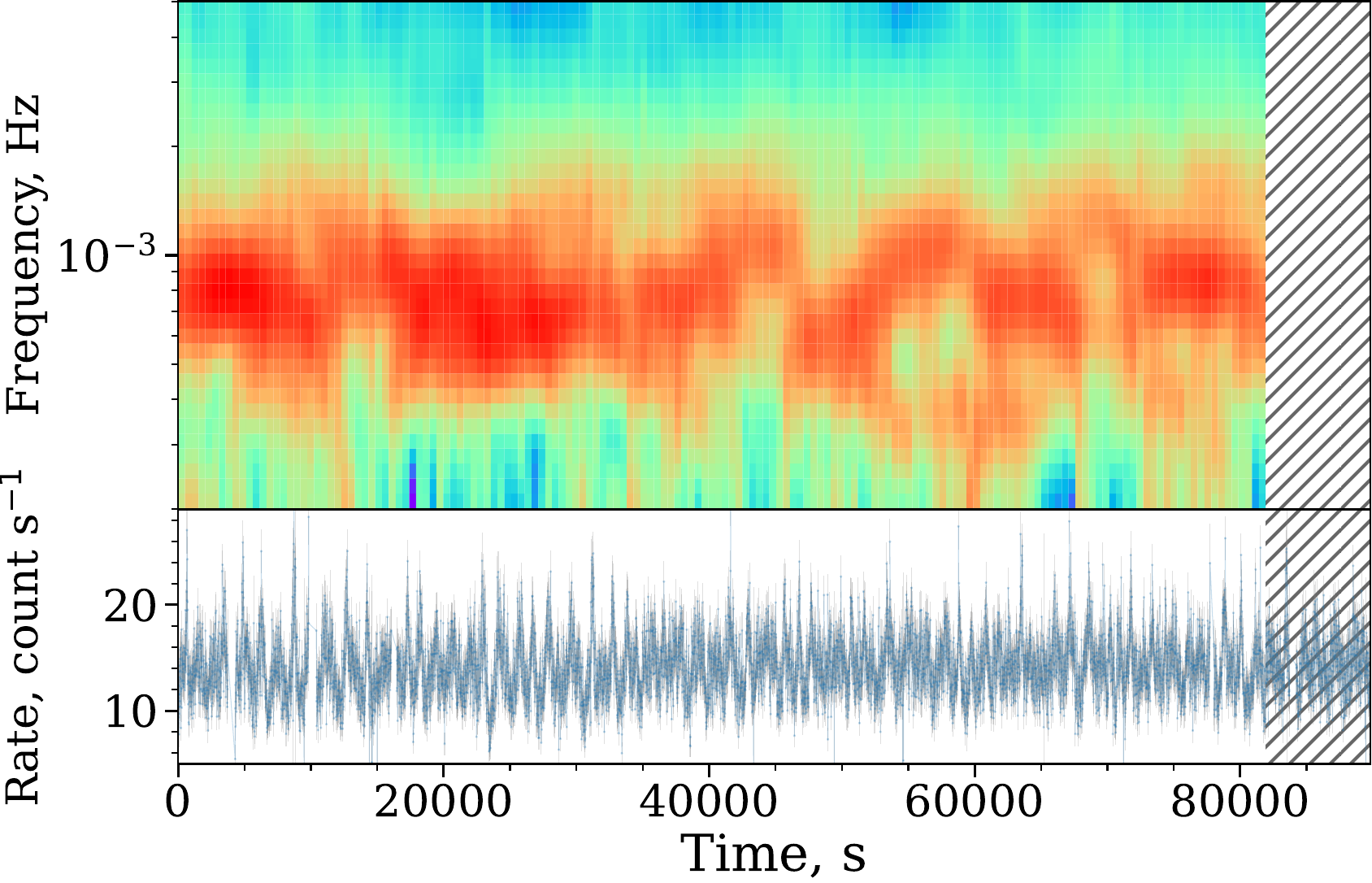}
\caption{Smoothed dynamic power spectrum (top) and light curve (bottom) of NuObs1 in the 3--79~keV energy band. The dynamic power spectrum is computed using a sliding window of 8192~s with a step of 512~s. The hatched region indicates the part of the time range where the sliding window cannot be applied without extending beyond the data boundaries.
The light curve is binned with a 10~s time resolution. }
\label{fig:qpo_evolution}
\end{figure}

Upon closer inspection, the light curve of \src\ reveals quasi-periodic variability with a characteristic timescale of $\sim10^3$~s (bottom panel of Fig.~\ref{fig:qpo_evolution}). To investigate this behavior, we computed a dynamic power density spectrum (PDS) from the NuObs1 barycenter-corrected 3--79~keV light curve using the Lomb-Scargle periodogram \citep{Lomb1976,Scargle1982}, as implemented in the \texttt{astropy} Python library. Each time bin in the dynamic PDS corresponds to a power spectrum computed over the following  8192-s interval, with a step size of 512~s. Each individual power spectrum was logarithmically rebinned in frequency to enhance visibility (top panel of Fig.~\ref{fig:qpo_evolution}). The quasi-periodic variability is most prominent during the first 32.8~ks of the observation (MJD~60797.8239--60798.2041). The hatched region on the right-hand side of Fig.~\ref{fig:qpo_evolution} marks the interval where the sliding window cannot be applied without extending beyond the observational data span.

To characterize the QPO, we extracted this segment of the light curve and computed an averaged PDS using the \texttt{powspec} tool from the \textsc{xronos} package. The PDS was calculated using root-mean-square (rms) normalization to express variability in terms of fractional rms amplitude. The white noise level was subtracted and the spectrum was  also logarithmically rebinned geometrically with a factor of 1.1. The same procedure is used for all other power spectra presented in this work unless stated otherwise.

\begin{figure}
\centering
\includegraphics[width=0.95\linewidth]{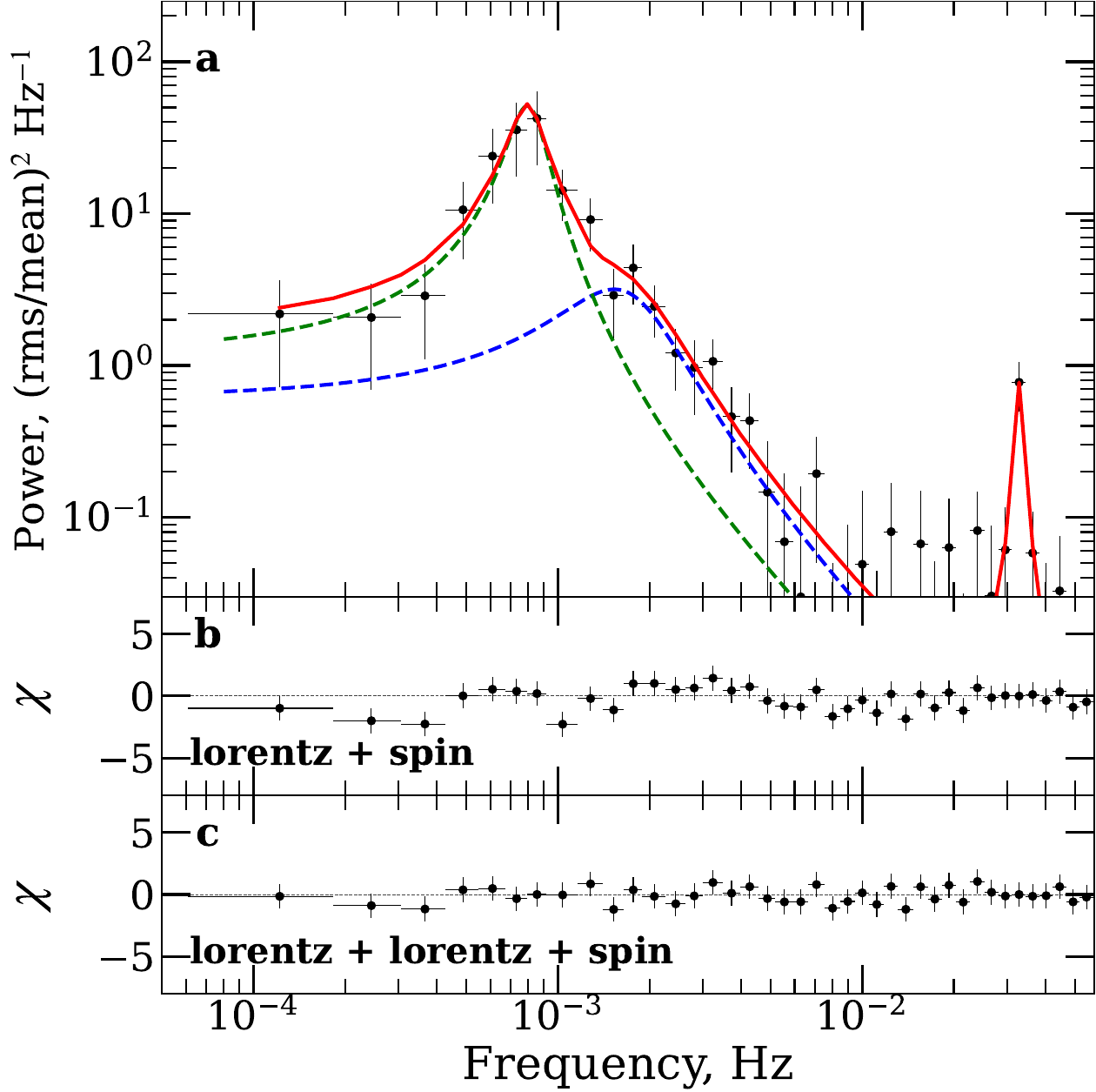}
\caption{Rms-normalized PDS of \src during the QPO-active interval for NuObs1 (3--79~keV) computed from the first 32.8~ks of the observation where the QPO is most prominent. The best-fit model is shown in red, with the green dashed line for the QPO component and the blue dashed line for the shoulder component. Panel (b) shows residuals to the model with only the QPO and spin frequency, while panel (c) shows residuals after adding the shoulder component.}
\label{fig:powerspec-nuobs1}
\end{figure}

The resulting PDS shown in Fig.~\ref{fig:powerspec-nuobs1} reveals a broad feature near $\sim10^{-3}$~Hz, consistent with a mHz QPO. This QPO is accompanied by an additional excess at higher frequencies, hereafter referred to as the ``shoulder.'' A similar QPO-shoulder structure, along with low-frequency red noise, was previously observed in the super-Eddington pulsar M51 ULX-7 \citep{Imbrogno2024}. Following the same modeling approach, we fit the PDS using components for the QPO, the shoulder, and the pulsar spin frequency.  We excluded the red noise component because the available data do not provide sufficient coverage at low frequencies (below $10^{-4}$~Hz) to constrain it reliably.

We converted the observed PDS into an \textsc{xspec}-compatible format using the \texttt{flx2xsp} utility and fitted it with a sum of three \texttt{lorentz} models in \textsc{xspec}, using a Whittle likelihood for the parameter estimation.
This approach is more suitable than least squares when dealing with exponentially distributed periodogram powers \citep[e.g.,][]{Whittle1953, Whittle1957}. All model parameters were left free during fitting, except for the spin frequency, which was fixed at $\nu_{\rm spin} = 1/30.44970$~Hz. The best-fit model yielded a QPO centered at $\nu_{\mathrm{QPO}} = 0.8\pm0.1$~mHz, with a quality factor of $Q = 3.2^{+10.6}_{-1.8}$ and a fractional rms amplitude of $14\pm5$\%. The shoulder component was located at $\nu_{\mathrm{shoulder}} \approx 1.5$~mHz, with a poorly constrained quality factor. The best-fit values and their uncertainties are listed in Table~\ref{tab:qpo_params}.

To assess the significance of the detected QPO, we performed Monte Carlo simulations using the \texttt{simftest}\footnote{\url{https://heasarc.gsfc.nasa.gov/xanadu/xspec/manual/node126.html}} procedure from the \textsc{xspec} package, providing the difference in Whittle statistic between the models with and without the QPO component. We ran $10^4$ simulations including the narrow Lorentzian QPO component. The resulting detection significance corresponds to at least 3$\sigma$. Thus, the QPO is firmly detected in the course of the NuObs1 observation, when the source was accreting at a bolometric luminosity of $L_{\rm bol} \approx 2.5 \times 10^{38}$~\lum\ (see Table~\ref{table:spec_params}).

\begin{figure}
\centering
\includegraphics[width=0.95\linewidth]{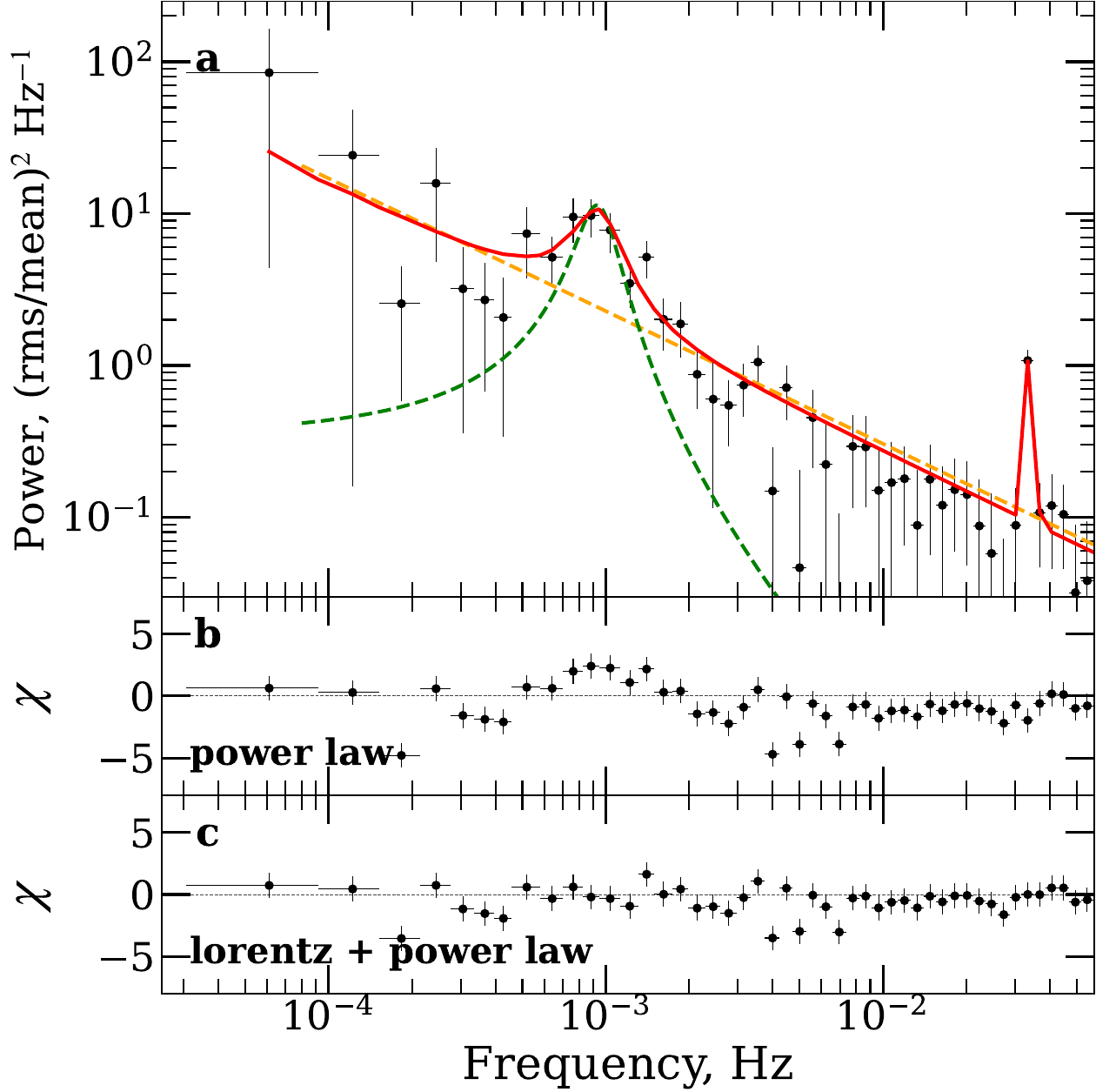}
\caption{Same as Fig.~\ref{fig:powerspec-nuobs1} but for ArtObs1 (4--25~keV) observation.  
Here the orange dashed line shows the red-noise component. 
Panel (b) shows the residuals for the model with the red noise only, while panel (c) shows residuals after adding the QPO component.}
\label{fig:powerspec-artobs1}
\end{figure}

\begin{table}
\caption{Best-fit parameters of the QPO and shoulder components in the PDS model. }
\label{tab:qpo_params}
\centering
\begin{tabular}{lcc}
\hline\hline
Parameter & NuObs1 & ArtObs1 \\
\hline
$\nu_\mathrm{QPO}$, mHz   & $0.80^{+0.11}_{-0.04}$ & $0.9\pm0.1$ \\
$Q_\mathrm{QPO}$     & $3.2^{+10.6}_{-1.8}$ &  $2.8^{+6.5}_{-1.7}$\\
QPO frac. rms, \% &  $14\pm5$  & $10^{+3}_{-1}$\\
$\nu_\mathrm{shoulder}$, mHz    & $1.5^{+1.9}_{-1.2}$ &  \\
$Q_\mathrm{shoulder}$      &  $2^{+20}_{-2}$ & \\
Shoulder frac. rms, \% &  $12^{+6}_{-2}$ & \\
\hline
\end{tabular}
\tablefoot{For NuObs1, the PDS was computed from the first 32.8~ks of the observation, where the QPO is most prominent.}
\end{table}

To verify the presence and stability of the QPO, we independently analyzed the 4--25~keV ArtObs1 data, obtained just a few days earlier (see Fig.~\ref{fig:powerspec-artobs1}). The ART-XC PDS also exhibits a QPO at approximately 0.9~mHz (see Table~\ref{tab:qpo_params}), with  parameters that are consistent with those observed in NuObs1. For \mbox{ArtObs1}, we included the red noise in the model as a simple power law and the shoulder component was not required for the fit.

\subsubsection{Pulse profiles}
\label{sec:timing}

A standard epoch-folding technique, implemented in the \texttt{efsearch} tool of the \textsc{ftools} package, was used to estimate the spin period values for the observations presented in this paper (see Table~\ref{table:periods} and the bottom panel of Fig.~\ref{fig:lightcurve}). The uncertainty was estimated from the distribution of measured periods obtained from a set of $10^3$ simulated light curves, following the procedure from \citet{Boldin2013}.

\begin{figure*}
    \centering
    \begin{subfigure}[t]{0.2\textwidth}
        \centering
        \includegraphics[width=0.98\linewidth]{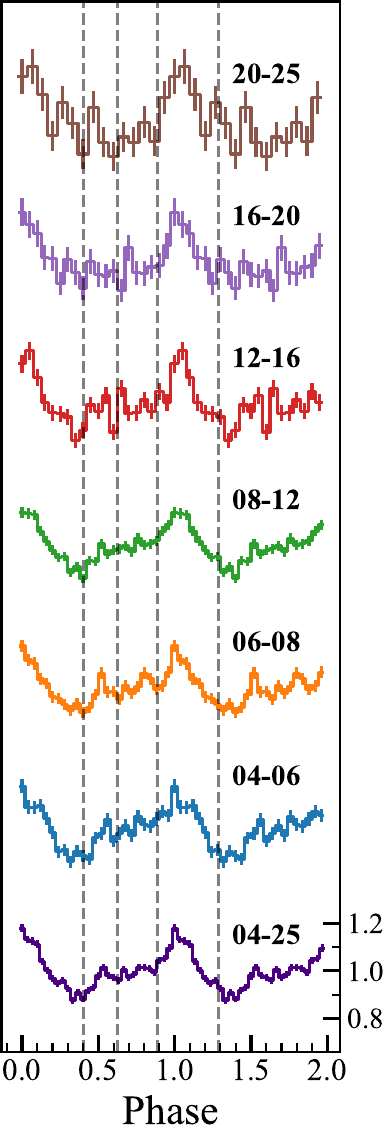}
        \caption{ArtObs1.}
        \label{fig:profiles_artobs1}
    \end{subfigure}%
    \hspace{0.01\textwidth}%
    \begin{subfigure}[t]{0.2\textwidth}  
        \centering
        \includegraphics[width=0.98\linewidth]{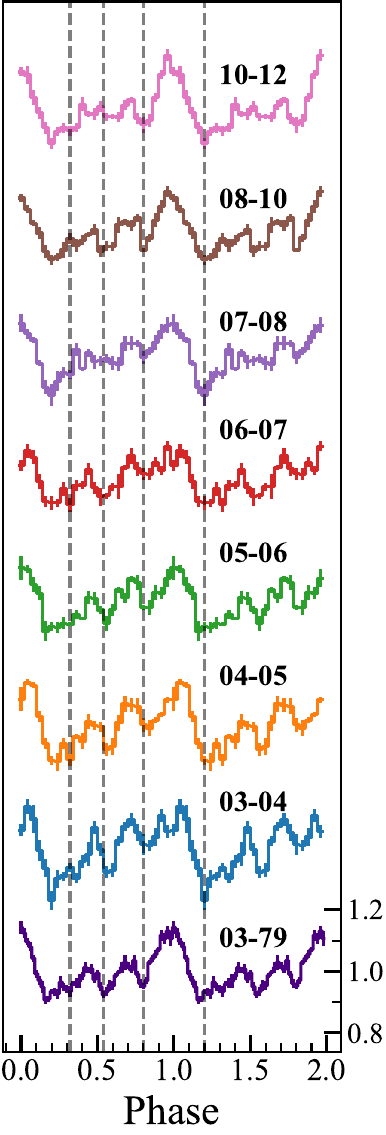}
                \caption{NuObs1.}

        \label{fig:profiles_nuobs1_1}
    \end{subfigure}
    \begin{subfigure}[t]{0.2\textwidth}  
        \includegraphics[width=0.98\linewidth]{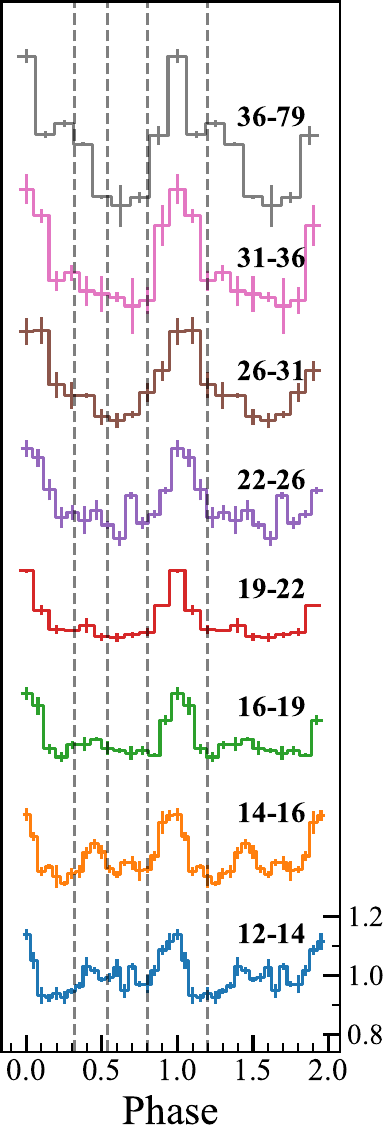}
        \caption{NuObs1 (cont.).}
        \label{fig:profiles_nuobs1_2}
    \end{subfigure}%
    \hspace{0.01\textwidth}%
    \begin{subfigure}[t]{0.2\textwidth}
        \centering
        \includegraphics[width=0.98\linewidth]{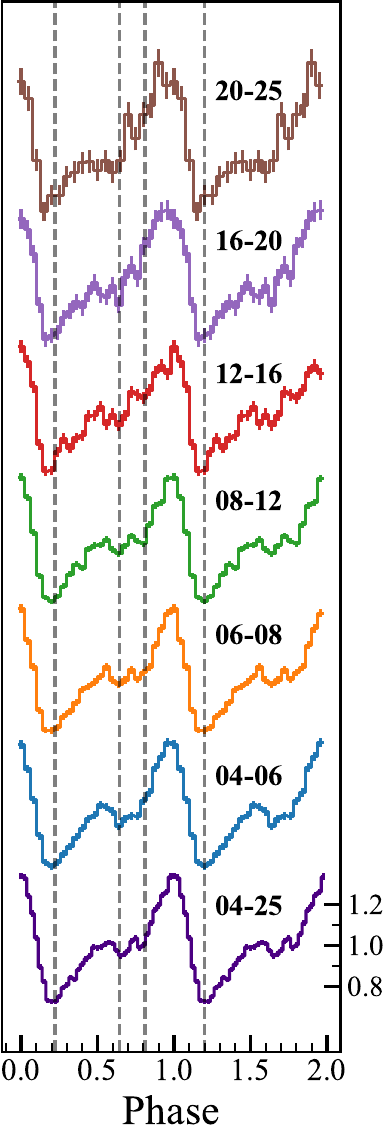}
        \caption{ArtObs2.}
        \label{fig:profiles_artobs2}
    \end{subfigure}\\[1.5ex] 

    \begin{subfigure}[t]{0.2\textwidth}  
        \centering
        \includegraphics[width=0.98\linewidth]{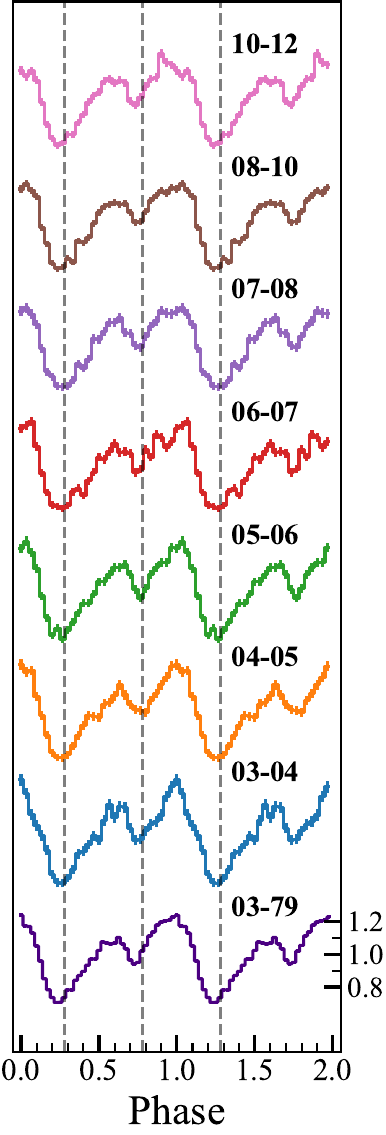}
        \caption{NuObs2.}
        \label{fig:profiles_nuobs2_1}
    \end{subfigure}
    \begin{subfigure}[t]{0.2\textwidth}  
        \includegraphics[width=0.98\linewidth]{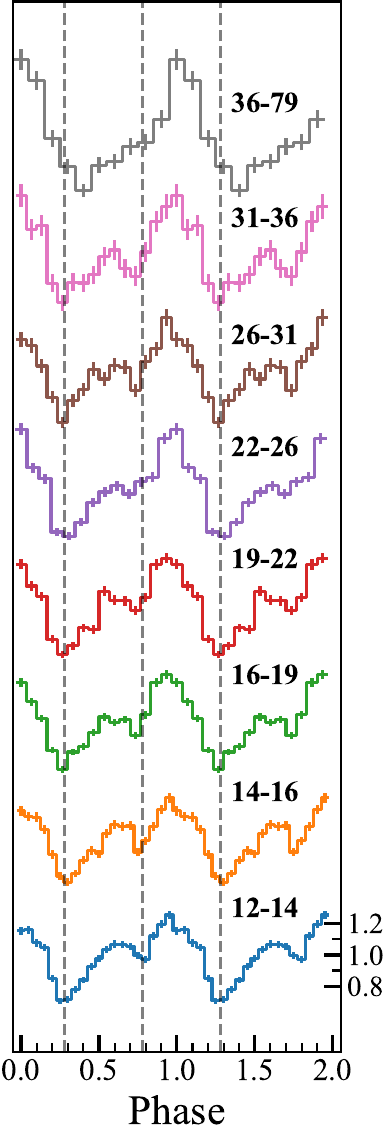}
        \caption{NuObs2 (cont.).}
        \label{fig:profiles_nuobs2_2}
    \end{subfigure}
    \hspace{0.01\textwidth}%
    \begin{subfigure}[t]{0.2\textwidth}
        \centering
        \includegraphics[width=0.98\linewidth]{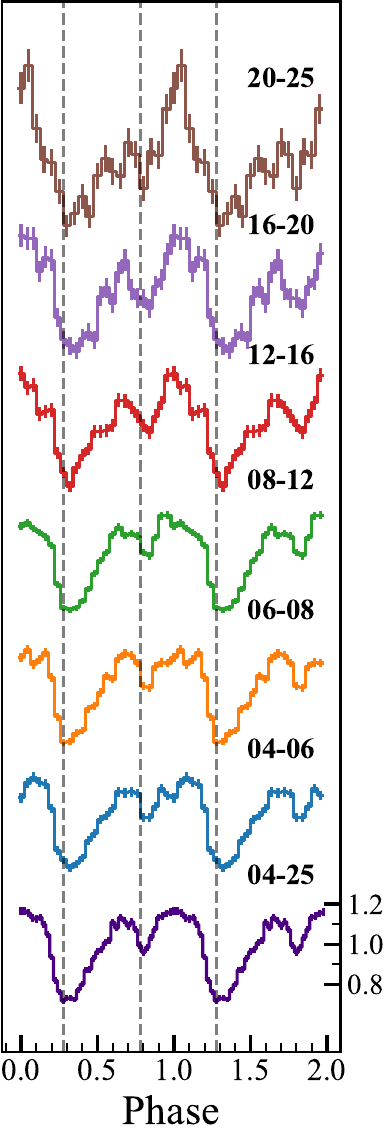}
        \caption{ArtObs3.}
        \label{fig:profiles_artobs3}
    \end{subfigure}
    \hspace{0.01\textwidth}%
    \begin{subfigure}[t]{0.2\textwidth}
        \centering
        \includegraphics[width=0.98\linewidth]{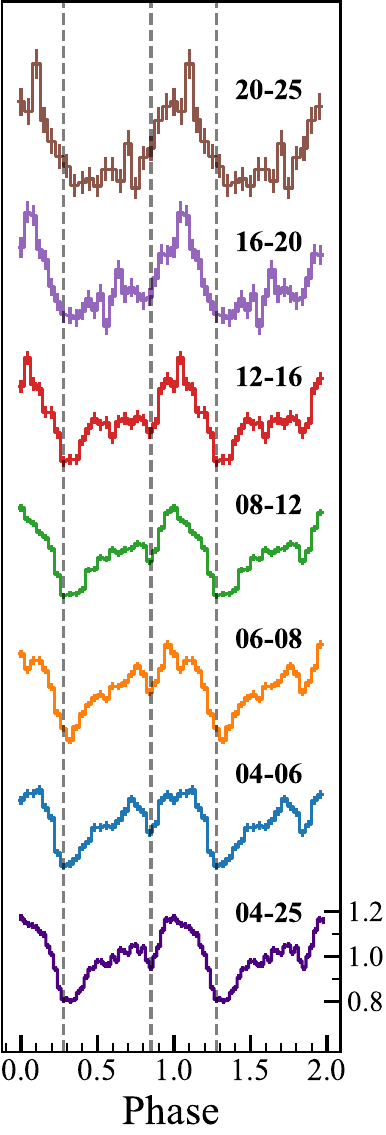}
        \caption{ArtObs4.}
        \label{fig:profiles_artobs4}
    \end{subfigure}%

\caption{Energy-resolved pulse profiles of \src. Gray dashed vertical lines mark the boundaries of the pulse peaks. The phase corresponding to the maximum of the profile is set to zero. Two full periods are shown per profile, each vertically offset for clarity. The flux in each energy band is normalized to its mean value. Energy ranges (in keV) are labeled next to each profile.}
\label{fig:combined_profiles}
\end{figure*}

Pulse profiles are a valuable diagnostic tool for investigating the emission characteristics of an NS and the properties of its surrounding environment. For \src, we generated pulse profiles by folding the barycenter-corrected light curves with the best-fit spin periods (see Table~\ref{table:periods}) using the \texttt{efold} task from the \textsc{xronos} package. The energy-resolved profiles for all epochs are presented in Fig.~\ref{fig:combined_profiles}.

The pulse profile during epoch~1 exhibits a three-peaked structure at low energies, which gradually evolves with increasing energy (see Figs.~\ref{fig:combined_profiles}a--c). As the energy increases, the secondary peaks weaken: the profile first transitions to a double-peaked shape and then to a single-peaked, nearly sinusoidal form above $\sim$26~keV. Notably, at these highest energies, a pronounced shoulder appears on the main peak around phases $\sim0.2$--0.5 (Figs.~\ref{fig:combined_profiles}c).

\begin{figure}
\centering
\includegraphics[width=0.95\columnwidth]{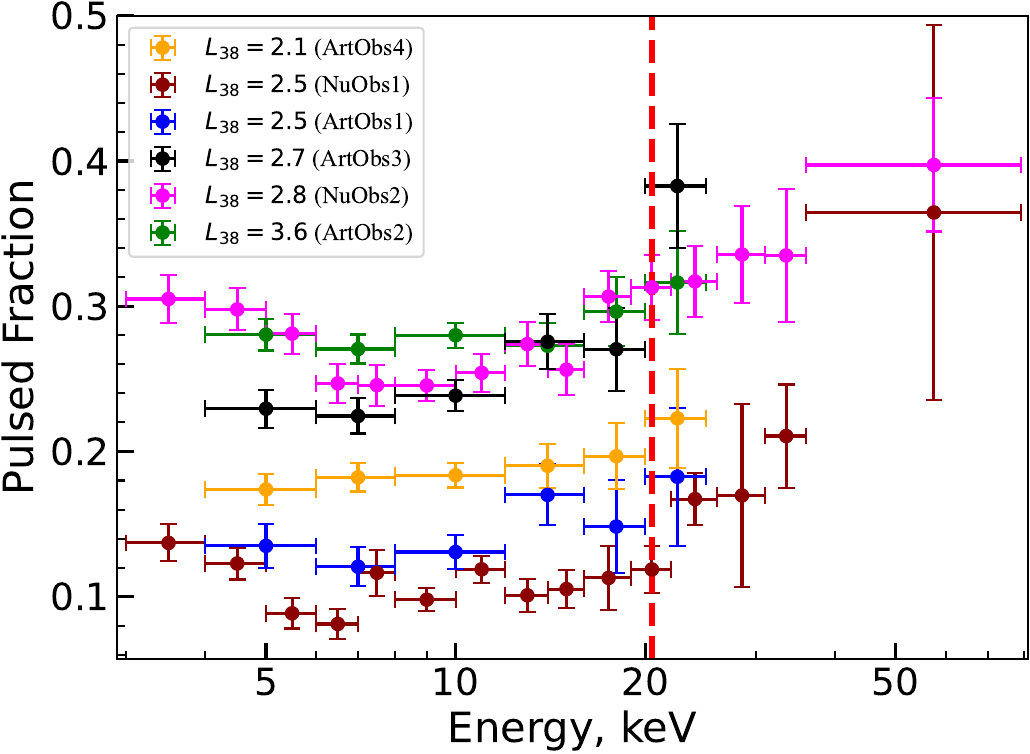}
\caption{Dependence of the PF on  energy. The dashed red line marks the energy above which secondary peaks disappear from the pulse profile in NuObs1, coinciding with the onset of a growth in the PF (see Sect.~\ref{sec:timing}). $L_{38}$ is the bolometric luminosity in units of $10^{38}$~\lum.}
\label{fig:pulsed}
\end{figure}

To explore these variations in more detail, we analyzed energy-resolved folded light curves to derive the energy dependence of the pulsed fraction (PF), which is commonly defined as [max(rate) -- min(rate)] / [max(rate) + min(rate)]. The PF, calculated using pulse profiles that were divided into 15 phase bins, displays a behavior that is unusual for XRPs (as shown in Fig.~\ref{fig:pulsed}). While most bright XRPs exhibit a monotonic increase in the PF with energy \citep{LutovinovTsygankov2009}, \src behaves differently: it remains at the level of $\approx10$\% up to 20 keV with a local minimum near the  6.4~keV iron line. At higher energies, PF begins to rise, reaching values of approximately 35\% in the 36--79~keV energy band. This increase in the PF coincides with the previously noted phase broadening of the main peak and the disappearance of the secondary peaks.

\begin{table*}
\centering
\caption{Spectral parameters for the best-fit model of \src\ based on data from epochs 1--5 .}
\begin{tabular}{lccccc} 
 \hline \hline
 \small{Parameter} & \multicolumn{4}{c}{\small{\texttt{cutoffpl+bbodyrad+bbodyrad+gauss}}}\\
  \hline
  & Epoch 1 & Epoch 2 & Epoch 3 & Epoch 4 & Epoch 5 \\
  \hline
const$_{\rm FPMA}$ & 1.000 (frozen) &  & 1.000 (frozen) &  &  \\
const$_{\rm FPMB}$ & $1.017\pm0.003$ &  & $1.028\pm0.003$  & &  \\
const$_{\rm ART-XC}$ & $0.925\pm0.003$ &  1.000 (frozen) & & 1.000 (frozen) & 1.000 (frozen) \\
const$_{\rm XRT,~PC}$ & $0.949^{+0.069}_{-0.066}$ &  &  & &  \\
const$_{\rm XRT,~WT}$ & $1.059\pm0.038$ &  $0.924^{+0.017}_{-0.023}$ & $1.067\pm0.020$& $0.896^{+0.015}_{-0.016}$& $0.976^{+0.019}_{-0.021}$ \\

$N_{\rm H}$, $10^{22}$~cm$^{-2}$ & 0.18 (frozen) & 0.18 (frozen) & 0.18 (frozen) & 0.18 (frozen) & 0.18 (frozen) \\
 Photon index ($\Gamma$) & $-0.55\pm0.04$ & $-0.7^{+0.3}_{-0.2}$ & $-0.33_{-0.04}^{+0.03}$ & $-0.1\pm0.1$ & $-0.2\pm0.1$\\
$E_{\rm fold}$, keV & $8.6\pm0.1$ & $8.3^{+1.6}_{-0.8}$ & $9.4^{+0.2}_{-0.1}$ & $12\pm1$ & $10\pm1$ \\
$T_{\rm bb}$, keV & $1.29\pm0.03$ & $1.35^{+0.04}_{-0.06}$ & $1.1\pm0.4$ & $1.29\pm0.03$  & $1.24^{+0.03}_{-0.04}$\\
$R_{\rm bb}$, km &  $7.4\pm0.2$ & $9.8^{+0.2}_{-0.3}$ & $10.2^{+0.6}_{-0.5}$ &  $7.6\pm0.4$ & $7.3\pm0.3$\\
$T_{\rm excess}$, keV & $0.22\pm0.02$ & 0.22 (frozen) & 0.22 (frozen) & 0.22 (frozen) & 0.22 (frozen)\\
$R_{\rm excess}$, km &  $164^{+38}_{-30}$ & $163^{+18}_{-17}$ & $167\pm4$ & $161\pm4$ & $141^{+4}_{-5}$\\
$E_{\rm iron}$, keV & 6.4 (frozen) & 6.4 (frozen) & 6.4 (frozen) & 6.4 (frozen) & 6.4 (frozen)\\
$\sigma_{\rm iron}$, keV & $0.35\pm0.07$&0.1 (frozen)&$0.5\pm0.1$& 0.1 (frozen) & 0.1 (frozen)\\
EW$_{\rm iron}$, keV & $0.08\pm0.01$&$0.05\pm0.01$&$0.11\pm0.01$ & $0.01\pm0.01$ & $0.05\pm0.01$\\
 
 Flux \tablefootmark{a} & $5.35\pm0.02$ & $7.8^{+0.3}_{-0.2}$ & $5.96\pm0.03$ & $5.7\pm0.1$ & $4.5\pm0.1$\\
 Luminosity\tablefootmark{b} &  $2.50\pm0.01$&$3.6\pm0.1$ & $2.78\pm0.01$  & $2.7\pm0.1$ & $2.1\pm0.1$\\
W-statistic/d.o.f. &  3725/3654 & 875/819 & 3387/3451 & 1029/903 &  941/849\\
 \hline
\end{tabular}
\tablefoot{
\tablefoottext{a}{Unabsorbed flux in the 0.1--100~keV range in units $10^{-10}$ \flux.}
\tablefoottext{b}{Unabsorbed luminosity in the 0.1--100~keV range in units $10^{38}$~\lum.}
}
\label{table:spec_params}
\end{table*}

\subsection{Timing analysis during QPO-free interval}
\label{sec:timing_noqpo} 

\begin{figure*}
\centering
\includegraphics[width=0.38\linewidth]{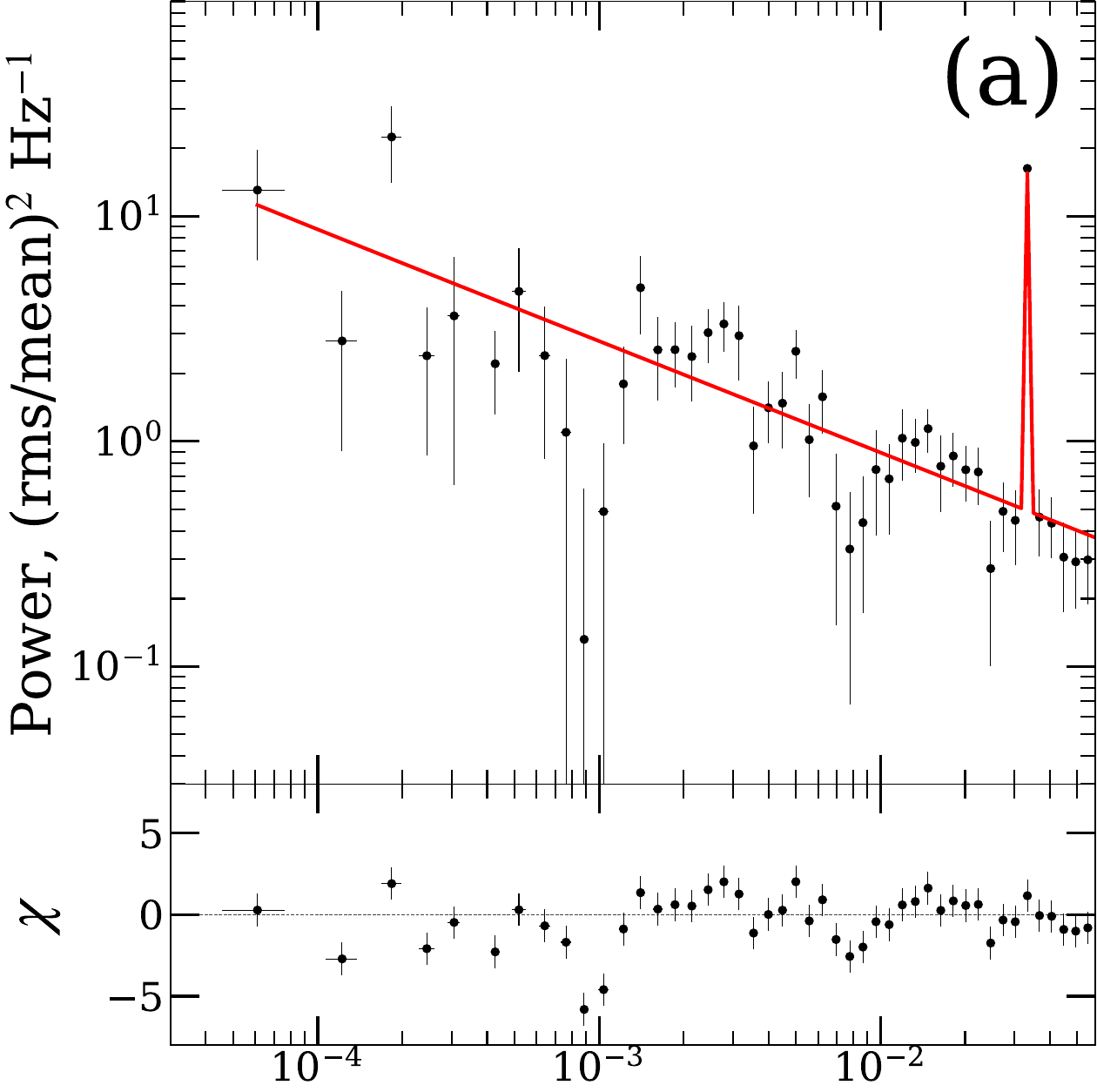}
\hspace*{1cm}
\includegraphics[width=0.38\linewidth]{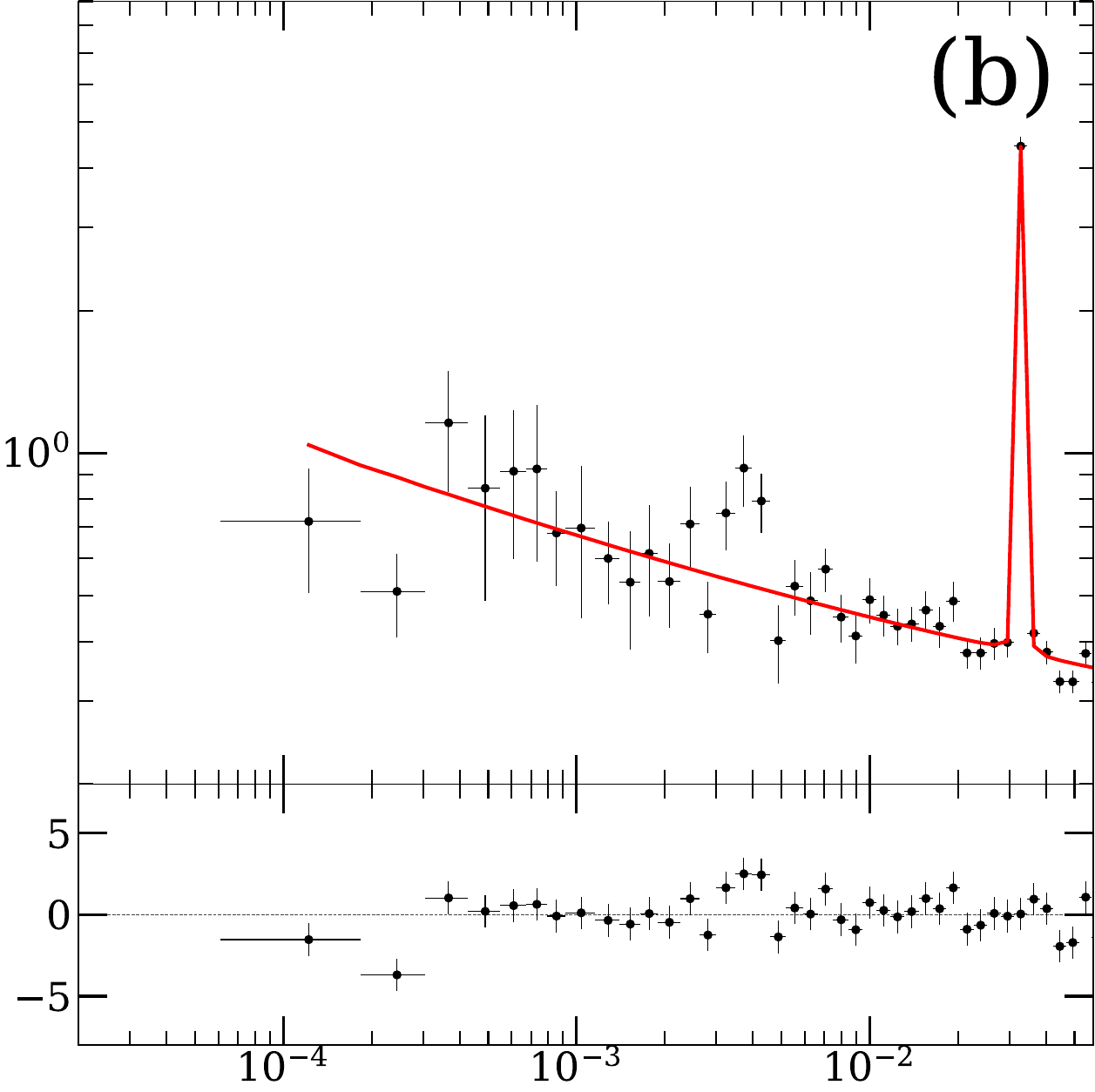} 

\vspace*{0.5cm}
\includegraphics[width=0.38\linewidth]{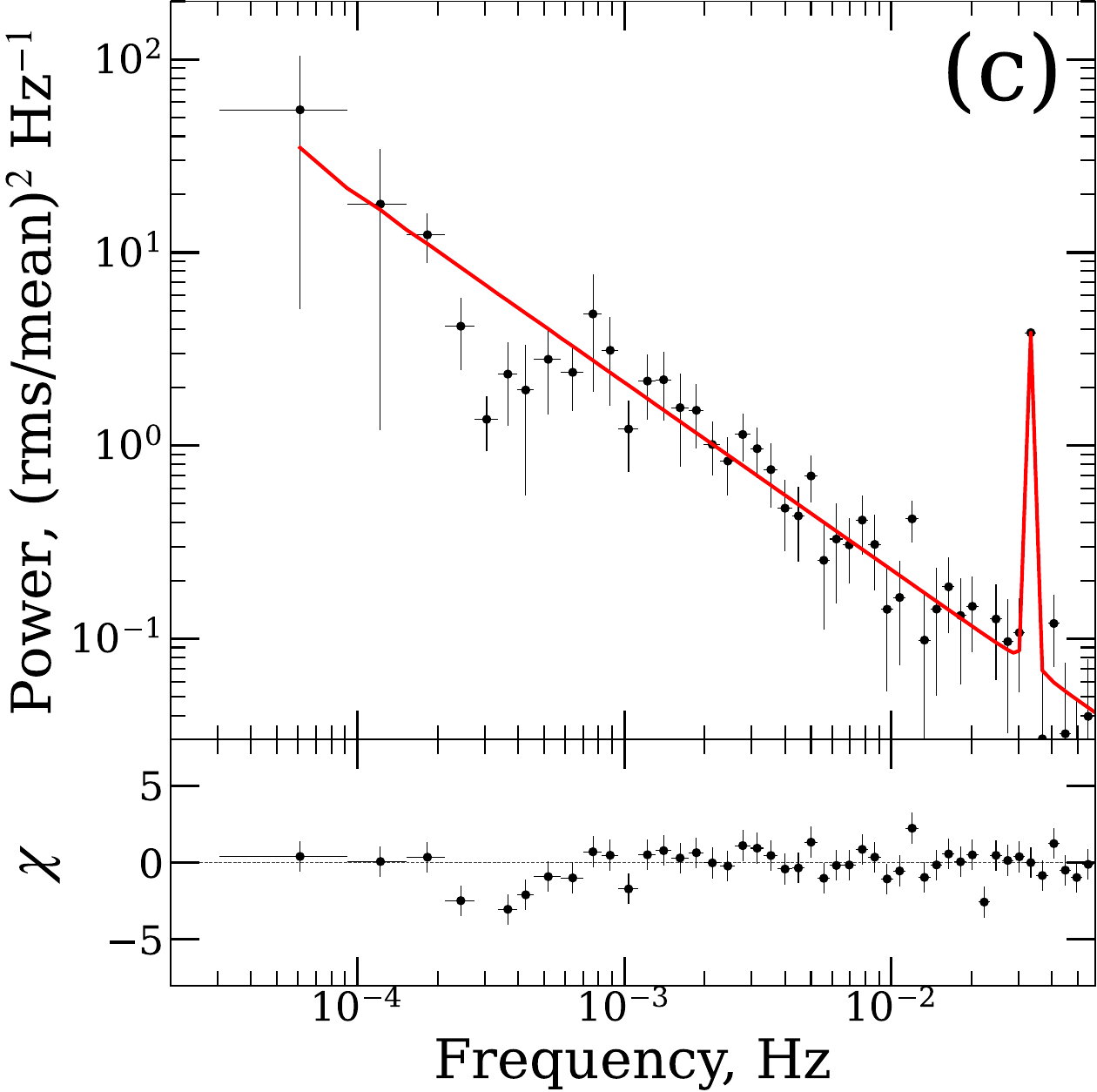}
\hspace*{1cm}
\includegraphics[width=0.38\linewidth]{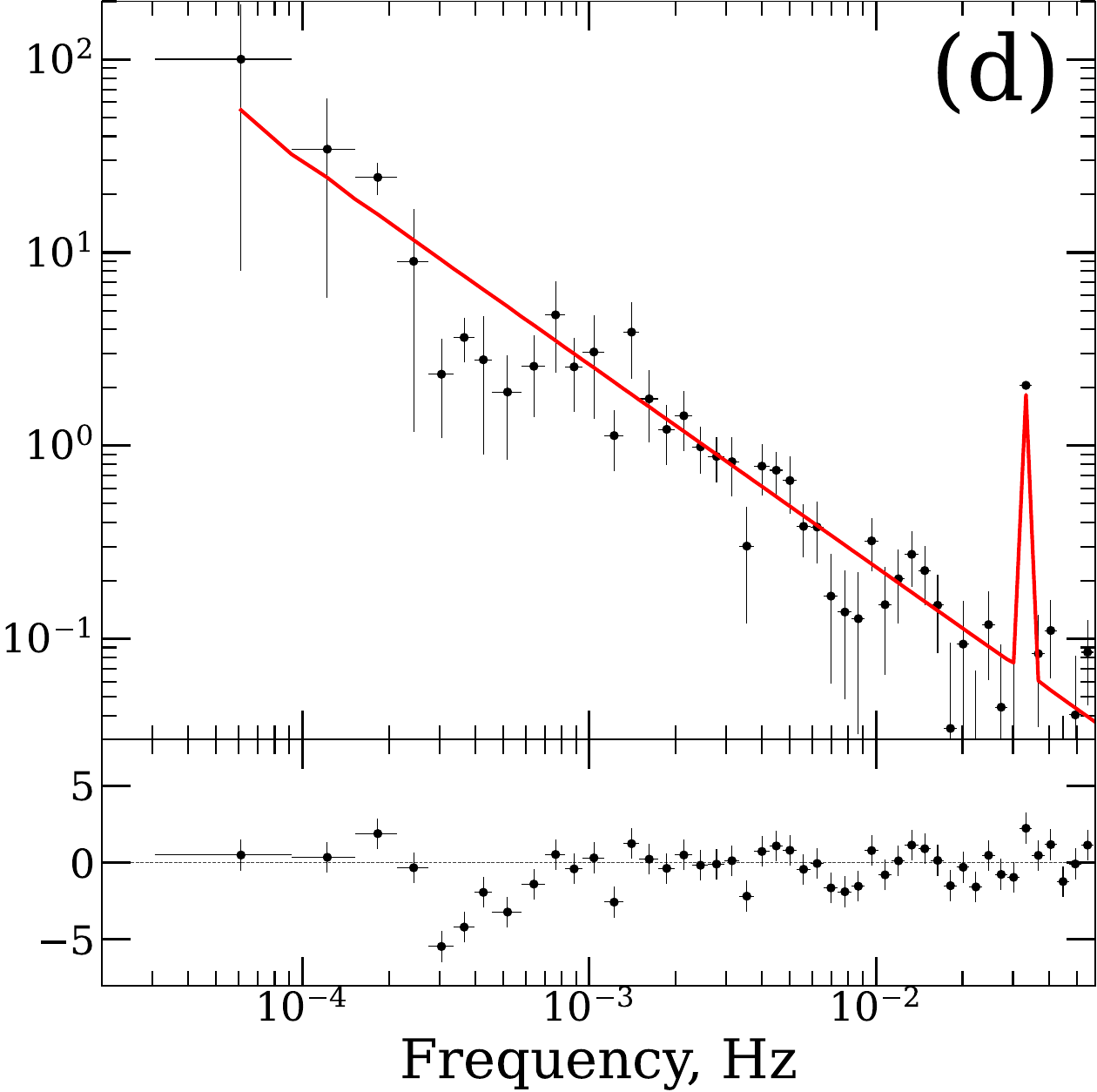}
\caption{Rms-normalized PDS of \src obtained after the initial detection. Upper subpanels show the data (black crosses) and the best-fit model (red line). 
The lower subpanels show the residuals. 
Panels (a), (b), (c), and (d) correspond to ArtObs2, NuObs2, ArtObs3, and ArtObs4 datasets, respectively.  
See text for details.}.
\label{fig:all_pds}
\end{figure*}

On MJD~60813, epoch~2 observations were performed at a higher luminosity of $L_{\rm bol} = 3.6\times10^{38}$~\lum (see Table~\ref{table:spec_params}), which is approximately 1.4 times higher than in epoch~1. One of the most notable results is the complete disappearance of the QPO from the PDS (see Fig.~\ref{fig:all_pds}a), despite improved statistics owing to the longer exposure and higher flux. Thus, the QPO vanished within 14 days of its last detection.
The PDS can be described as a red noise (a simple power law) and the QPO component was not required. However, this transition is accompanied by a sharp increase in PF in the 4--25~keV band, rising from an average value of $\approx 10$\% in epoch 1 to $\approx 26$\% in epoch 2 (see Fig.~\ref{fig:pulsed}).

An energy-resolved analysis of the pulse profile reveals a similar triple-peaked shape at all energies up to 25~keV (Fig.~\ref{fig:profiles_artobs2}). However, this structure appears significantly distorted compared to the previously observed pattern at epoch 1, despite only a modest increase in the flux. The relative phase spans of the peaks have changed substantially, indicating a change in the pulse morphology. The associated increase in PF, which remains at $\sim26$\% in all energy ranges up to 25 keV (Fig.~\ref{fig:pulsed}), along with the altered pulse shape, suggests a change in the beam configuration.

Subsequently, observations of epochs 3 and 4 were performed at a bolometric luminosity of $L_{\rm bol} =$ (2.7--2.8)$\times10^{38}$~\lum (see Sect.~\ref{sec:spectrum}), which is very similar to that of epoch~1. Despite that, no evidence of the previously detected QPO was found in the PDS (Figs.~\ref{fig:all_pds}b,c) and the timing properties differed markedly from those observed in epoch 1 (see Fig.~\ref{fig:combined_profiles}e,f).  The PF remains at approximately 25\% up to 20~keV and then gradually increases to about 40\% at higher energies (Fig.~\ref{fig:pulsed}). The 3--79~keV pulse profile of \src\ exhibits a double-peaked structure (Figs.~\ref{fig:combined_profiles}e--g). An exception is seen in the 36--79~keV range, where the profile becomes distinctly single-peaked. 

Finally, epoch~5 is of particular interest, as it is the only observation obtained at a luminosity lower than that of epoch~1, where the QPO was detected (with epoch~5 reaching $L_{\rm bol} = 2.1 \times 10^{38}$~\lum), yet it still shows no evidence of the QPO (Fig.~\ref{fig:all_pds}d). The energy-resolved pulse profile retains a double-peaked structure at lower energies (see Fig.~\ref{fig:combined_profiles}h); the second peak vanishes above 16~keV. The PF remains moderately higher than in epoch~1, though lower than in the other QPO-free states (epochs 2 to 4), reaching 17\% in the 4--25~keV range.

\subsection{Spectral analysis}
\label{sec:spectrum}

We performed the pulse phase-averaged spectral analysis of \src\ using data from epochs 1--5 in order to trace the evolution of the spectral parameters in different states. We excluded the 3--4~keV range from the  \nustar data analysis due to cross-calibration difficulties with \swift, and energies below 0.8 and 0.5 keV were ignored for \swift in WT and PC modes, respectively.

\begin{figure}
\centering
\includegraphics[width=0.85\linewidth]{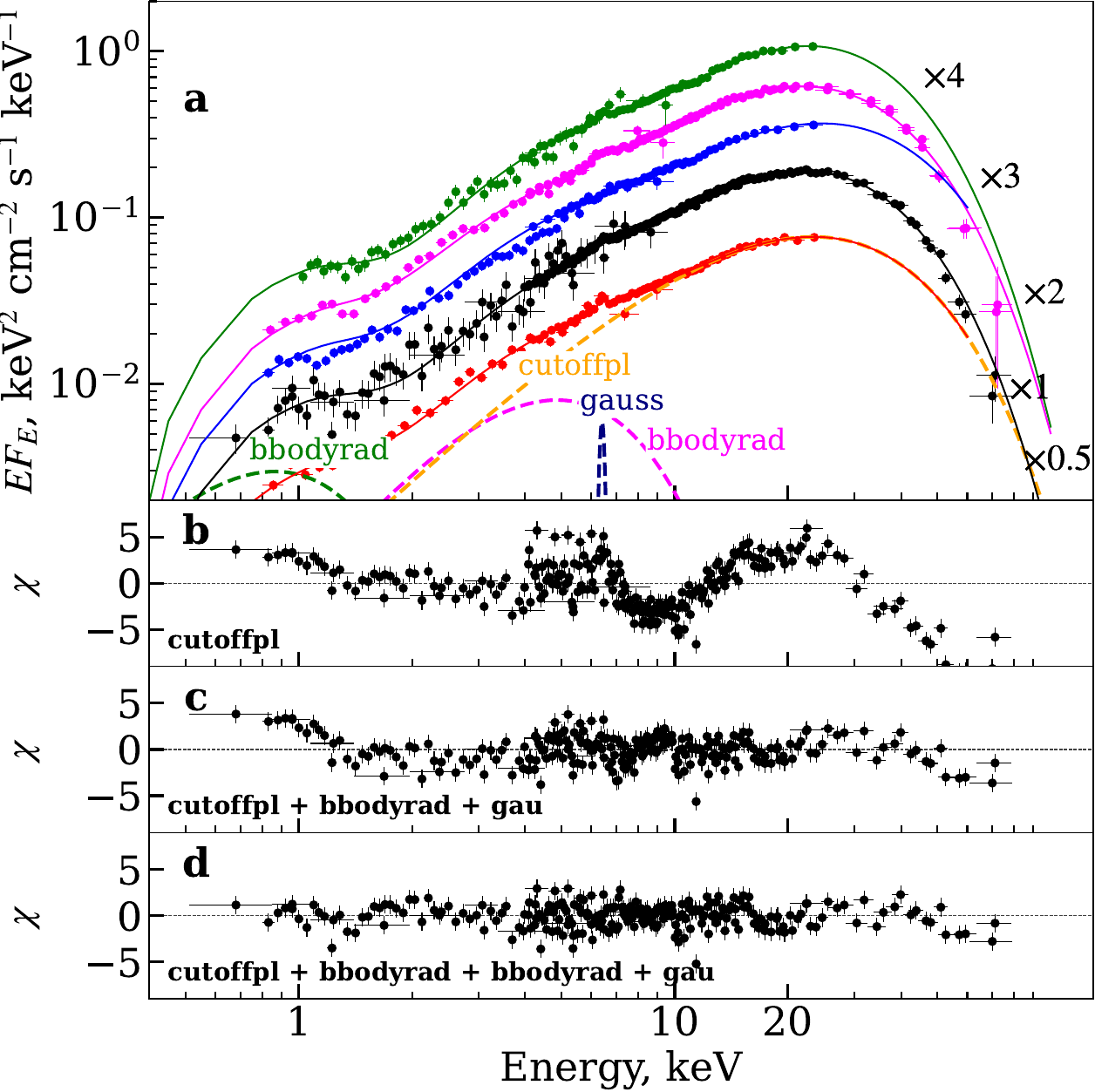}
\caption{
Unfolded energy spectra of \src. The spectra were spaced along the y-axis for visual convenience with the factors shown.
The black, green, magenta, blue, and red points correspond to the epoch 1--5 data and their respective model approximations.
The dashed lines represent the individual model components (\texttt{cutoffpl}, \texttt{bbodyrad}, \texttt{gauss}, and a second \texttt{bbodyrad}) for the epoch~5 dataset.
Panels (b), (c), and (d) show the residuals of the epoch~1 dataset when fitted with the \texttt{cutoffpl} model alone, with an additional \texttt{bbodyrad} component, and with two \texttt{bbodyrad} components, respectively.
}
\label{fig:spectrum}
\end{figure}

To approximate the source spectrum in epoch 1, we adopted the canonical accreting XRP model \citep[e.g.,][]{Filippova2005,Coburn2002}, consisting of a power-law continuum with an exponential cutoff at high energies. However, a simple \texttt{cutoffpl} model fails to reproduce the continuum adequately, leaving structured residuals (see Fig.~\ref{fig:spectrum}b). To improve the fit, we added a blackbody component with a temperature of $T_{\rm bb} \approx 1.3$~keV, motivated by a similar approach used in modeling other super-Eddington pulsars, such as Swift\,J0243.6+6124 \citep{Tao2019, Bykov2022} and SMC X-3 \citep{Pottschmidt2016,Tsygankov2017b}, where comparable blackbody temperatures were found. In our case, this significantly improved the fit (W-stat./d.o.f. = 1.08, see Fig.~\ref{fig:spectrum}c). Furthermore, after taking into account the soft excess with $T_{\rm excess} = 0.22$~keV and $R_{\rm excess}=160$~km, previously reported in \citet{Yokogawa2000}, we were able to reduce W-stat./d.o.f. ratio to the value of 1.01 (Fig.~\ref{fig:spectrum}d). The inclusion of the hot \texttt{bbodyrad} component is also required when adopting an alternative continuum model such as \texttt{po$\times$highecut} \citep{Coburn2002}, reinforcing its necessity. We were unable to detect the presence of any cyclotron absorption features. The best-fit model parameters are presented in Table~\ref{table:spec_params}.

The spectrum also shows an iron fluorescence line at 6.4~keV, previously reported by \citet{Yokogawa2000}, which is modeled with a narrow Gaussian component \texttt{gauss}.  The line width $\sigma$ was fixed at 0.1~keV for the \srg/ART-XC data, while it was left as a free parameter for the \nustar observations. A cross-calibration constant is included to account for possible normalization differences between the FPMA, FPMB, ART-XC, and \swift/XRT. 

To investigate the changes in the spectral properties, we analyzed the epoch~2 spectra using the same model as in epoch~1, which remained statistically preferred. For epochs~2--5, the soft excess temperature was fixed at $T_{\rm excess} = 0.22$~keV to reduce parameter degeneracy, adopting the best-fit value from epoch~1 (see Table~\ref{table:spec_params}). 

Although ArtObs2 is separated by approximately five days from the closest available \swift/XRT observation (ObsID 00019718011; see Table~\ref{table:all_obs}), both datasets exhibit consistent flux levels in the 4--10~keV band, with values of $1.4 \times 10^{-10}$~\flux. This agreement justifies their joint spectral fitting, which enables tighter constraints on spectral parameters. The spectrum was approximated using the same model as in the earlier analysis, and the fit remains statistically acceptable, with a W-stat./d.o.f. ratio of approximately 1.07. We also analyzed the spectra from epochs~3, 4, and 5 using the same model, which provided statistically acceptable fits with W-stat./d.o.f. values of 0.98, 1.14, and 1.11, respectively (Table~\ref{table:spec_params}). None of the spectra from any epoch show evidence of CRSFs in the 5--50~keV energy range.

\begin{figure*}
\centering
\includegraphics[width=0.7\columnwidth]{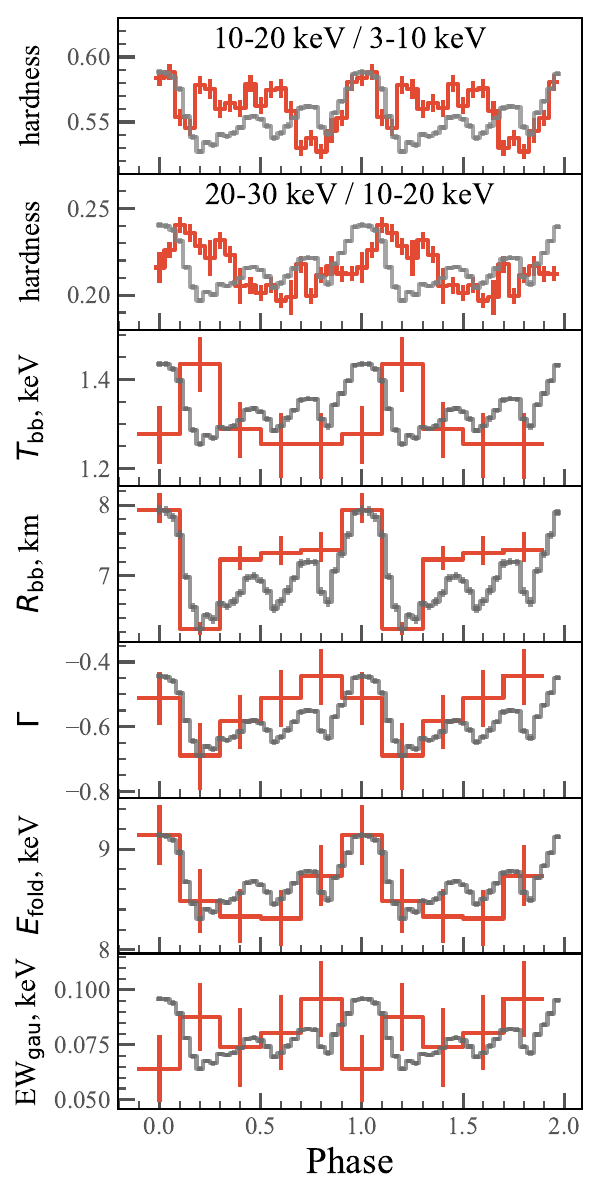} 
\includegraphics[width=0.7\columnwidth]{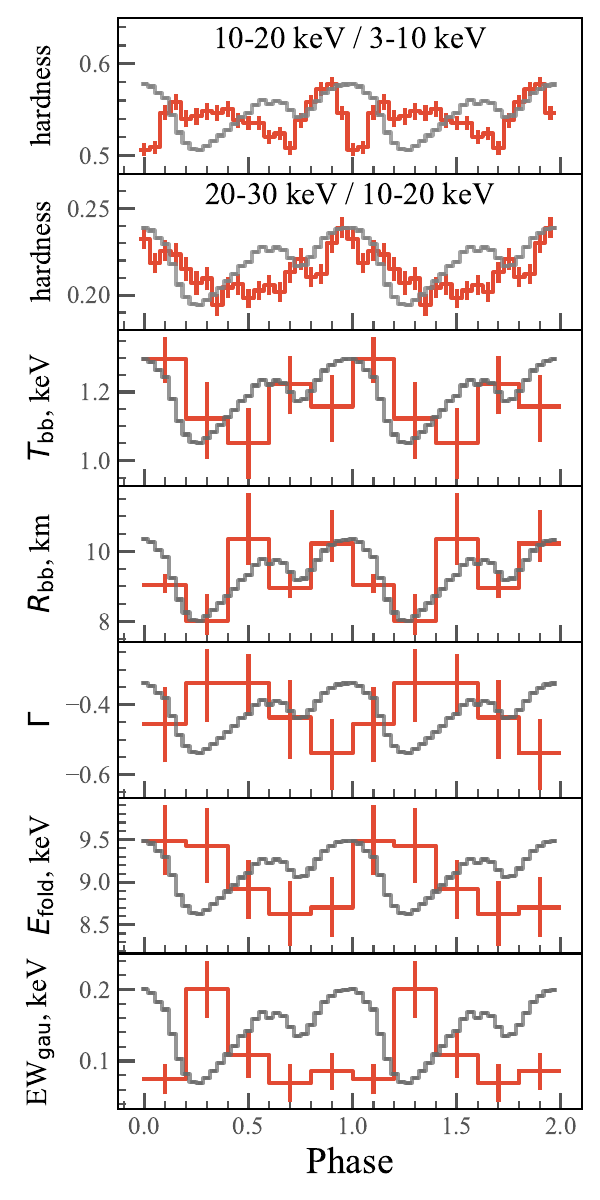} 
\caption{Dependence of the hardness ratios and the spectral parameters on the pulse phase for the NuObs1  (left) and NuObs2 (right). 
The averaged pulse profile in a wide energy range 3--79~keV is superimposed in gray for visual comparison.}
  \label{fig:phres_spec}
\end{figure*}

We also analyzed pulse-phase-resolved spectra for NuObs1 and NuObs2, obtained by dividing the data into five equally spaced pulse phase bins. Each spectrum was fitted using the same model as the phase-averaged spectrum except soft excess: \texttt{const $\times$ tbabs $\times$ (cutoffpl + bbodyrad + gauss)}. \swift/XRT data were not included. The energy of the iron line was frozen to the value $E_{\rm iron} = 6.4$~keV and the width values were frozen to $\sigma_{\rm iron}=0.35$ and 0.5~keV for NuObs1 and NuObs2, respectively, based on the best-fit values from the phase-averaged spectral analysis (see Table~\ref{table:spec_params}). The results can be found in Fig.~\ref{fig:phres_spec}. We were unable to detect any cyclotron absorption features in any of the spectra. Due to the low counting statistics and complexity of the spectral shape, the significance of the variations in the spectral parameters was hard to assess. 

\section{Discussion and conclusions}

The \nustar, \swift/XRT, and \srg/ART-XC observations of \src\ during its 2025 type~II outburst provide the first comprehensive broadband spectral and timing characterization of this XRP. The data span the 0.5--79~keV energy range and enable a detailed investigation of both pulse-phase-averaged and pulse-phase-resolved properties. The source exceeded the Eddington limit of $1.8\times10^{38}$~\lum for a canonical $1.4\msun$ NS during the outburst. This firmly places \src\ in the super-Eddington accretion regime, among the most luminous outbursts ever observed in BeXRB systems.

Throughout all broadband observations, the phase-averaged X-ray spectrum was characterized by a hard power-law continuum, modified by a high-energy exponential cutoff. The inclusion of two blackbody components with $T \approx 1.3$~keV, $R \approx 7$~km, and a cooler excess component with $T_{\rm excess} \approx 0.2$~keV, $R \approx 140-170$~km was required to accurately reproduce the observed spectrum. A narrow Fe~K$\alpha$ emission line was detected at 6.4~keV, with an equivalent width not exceeding $\approx110$~eV. No CRSF, a direct probe of the NS’s magnetic field \citep{Staubert2019}, was detected in either phase-averaged or phase-resolved  spectra of \src\ across the 5--50 keV range in any of the observations presented in this study, suggesting magnetic field strength $B < 5 \times 10^{11}$~G or $> 5 \times 10^{12}$~G. A shallow or complex-shaped feature below the detection threshold cannot be entirely ruled out.

We discovered the presence of a 0.8-mHz QPO in \src\ at a luminosity of $L_{\rm bol} \approx 2.5 \times 10^{38}$~\lum in the epoch~1 observations. It establishes \src as the fourth known super-Eddington XRP to exhibit mHz low-frequency quasi-periodic variability \citep[for a review, see][]{Cemeljic2025, Veresvarska2025}, and the first such detection in a super-Eddington luminosity state outside the class of confirmed ultraluminous X-ray pulsars (ULXPs). This newly detected QPO does not match the 1.27~Hz feature previously reported   by \citet{Kaur2007}, which was observed during an earlier outburst at a similar luminosity level of $\sim10^{38}$~\lum. 

In NuObs1, the energy-resolved pulse profile of \src\ evolved from a complex three-peaked structure at soft X-rays to a single, nearly sinusoidal shape above 26~keV (Fig.~\ref{fig:combined_profiles}b,c).
At low energies ($\lesssim$10~keV), the PF in \src\ remains relatively low, at around $\sim10$\%, consistent with values observed in other super-Eddington XRPs exhibiting mHz QPOs. For example,  M51~ULX-7 ($P_{\rm spin}\approx$2.8~s, $\nu_{\rm QPO}\approx$0.5--0.6~mHz; \citealt{Imbrogno2024}) shows an upper limit of 7\%, NGC~7793~P13 ($P_{\rm spin}\approx$0.4~s, $\nu_{\rm QPO}\approx10$~mHz; \citealt{Imbrogno2024a, Israel2025}) exhibits PFs in the range 5--10\%, and in case of M82~X-2 ($P_{\rm spin}\approx$ 1.3~s, $\nu_{\rm QPO}\approx$2.8--4.0~mHz; \citealt{Feng2010}), no detectable pulsations were reported during the only known QPO episode. All of these three systems had their QPOs observed at luminosities $L_{\rm X} \gtrsim 10^{39}$~\lum.

However, unlike these sources, \src\ was also observed at higher energies, where it exhibits a marked increase in the PF, rising to $\sim35$\% above 30~keV (see Sect.~\ref{sec:timing}). Since the other systems lack observations above 10~keV, any potential increase in PF at higher energies may have gone unnoticed due to a limited energy coverage. This suggests that hard X-ray observations could help identify NSs in ULXs showing mHz QPOs, even in the absence of low-energy pulsations.

In the subsequent  epoch 2 observations at the luminosity of $L_{\rm bol} \approx 3.6 \times 10^{38}$~\lum (see Table~\ref{table:spec_params}), the QPO was no longer detected. At the same time, the PF in the 4--25~keV energy band increased sharply from approximately 10\% to 26\%. The pulse-profile shape also exhibited a three-peaked structure, although the overall morphology appeared distorted. 

In  epochs~3 and 4, observations at the luminosities  $L_{\rm bol} \approx$(2.7--2.8)$\times 10^{38}$~\lum  indicate that the QPO was once again not detected and the PF showed similar to the epoch 1 behavior, rising to 40\% above 25~keV. Energy-resolved pulse profiles evolved into a two-peaked shape, transitioning to a sinusoidal, single-peaked profile above 36~keV. 

In epoch~5, at a lower luminosity of $L_{\rm bol} = 2.1 \times 10^{38}$\lum, no QPO was detected. The pulse profile was double-peaked at low energies, with the second peak vanishing above 16~keV, and the PF reached 17\% in the 4--25~keV range, somewhat below other QPO-free epochs. 

These findings indicate that the QPO exhibits a distinctly transient nature, appearing only under specific physical conditions, since it is absent at both higher and lower luminosities after the detection. Moreover, QPO presence is associated with a marked decrease in the PF.

The drastic frequency shift of more than three orders of magnitude between the 1.27~Hz QPO reported by \citet{Kaur2007} and the 0.8~mHz QPO detected in this work cannot be explained by standard QPO models such as the BFM \citep{Alpar1985} or the Keplerian frequency model \citep{vanderKlis1987}, which predict $\nu_{\rm QPO} \propto \dot{M}^{3/7}$. Explaining such a shift would require an unobserved change in the accretion rate by nearly seven orders of magnitude. This strongly suggests a different physical origin for the newly detected mHz QPO.

One explanation comes from the disk precession model proposed for ULXPs by \citet{Cemeljic2025}, in which global, rigid-body-like precession of a tilted inner accretion disk is driven by magnetic torques from the NS's inclined dipole. A key feature of this model is the inverse scaling of the QPO frequency with a spin period, $\nu_{\rm QPO} \propto 1/P_{\rm spin}$, which is largely independent of magnetic field strength or accretion rate, and primarily governed by the disk geometry. Specifically, the model requires the presence of a geometrically thick inner accretion disk capable of supporting global precession.

For SXP31.0, with a spin period of 30.45 s, the model predicts a QPO frequency of about 0.1 mHz, placing the observed 0.8 mHz signal within a factor of eight of the expected value. Considering the model’s sensitivity to uncertain geometric parameters, this discrepancy is not necessarily problematic. The proportionality constant, $K$, in the $\nu_{\rm QPO} = K/P_{\rm spin}$ relation depends sensitively on the disk geometry and may deviate from the  established $K \approx 3$~mHz\,s if the accretion flow in \src\ is not fully radiation-pressure dominated or geometrically thick, as assumed for confirmed ULXPs. Indeed, \src\ resides near the Eddington limit and may not sustain a thick, radiation-pressure-dominated disk typical of confirmed ULXPs \citep{Mushtukov2019}. Instead, it may represent a transitional regime, where the disk is only moderately inflated and geometrically intermediate. In such a configuration, the conditions required for coherent global precession may be only marginally satisfied, potentially shifting the QPO frequency away from the expected scaling. 

Alternatively, the observed 0.8-mHz QPO in \src\ may be interpreted within the magnetically driven precession (MDP) framework originally developed by \citet{2002ApJ...564..361S,2002ApJ...565.1134S}, a more general version of the global precession scenario later applied by \citet{Cemeljic2025}.
While the latter relies on a simplified scaling for a radiation-pressure-supported thick disk in spin equilibrium, the MDP model derives the precession frequency from first principles, balancing magnetic torques and viscous stresses in a warped accretion flow. Although the model in principle allows one to constrain the dipolar magnetic field, $B$, we avoid quoting specific values due to significant uncertainties in key parameters such as the viscosity parameter, $\alpha$, magnetic obliquity. $\theta$, and the disk thickness factor, $D(r)$. Importantly, the MDP framework of \citet{Cemeljic2025} was calibrated specifically for the ULXP systems, which display mHz QPO, which is an order of magnitude brighter than \src.

The MDP-based interpretation supports the scenario in which mHz QPOs originate in regions of the accretion flow that also influence the formation or visibility of pulsations. It also underscores the need for caution when using QPO frequencies in ULXs as diagnostics of the compact object mass, since the physical origin of the QPO may not be directly tied to Keplerian timescales. Finally, the apparent anticorrelation between QPOs and PF may partly explain the difficulty in detecting pulsations in many ULXs and suggests that the fraction of NSs among the ULX population could be higher than currently inferred.

Moreover, the current data provide the first observational constraint on how quickly the disappearance of the mHz QPO and the accompanying rise in PF can occur. In \src, this transition took place within a $\lesssim 14$~day interval between May~3 and May~18, setting an upper limit on the temporal evolution of this phenomenon in the super-Eddington accretion regime.

\begin{acknowledgements}
AS acknowledges support from the EDUFI Fellowship and Jenny and Antti Wihuri Foundation (grant no. 00240331). 
SST acknowledges financial support from the Centre for Astrophysics and Relativity at Dublin City University during his visit. 
AAM acknowledges UKRI Stephen Hawking fellowship. SVM, IYL, AAL and AYT acknowledge support from the Ministry of Science and Higher Education of RF grant 075-15-2024-647. We thank the anonymous referee for their careful reading of the paper and thoughtful suggestions, which helped improve the clarity of the results. 
AS also thanks Vadim Kravtsov for helpful discussions during the preparation of this work. 
This work is partially based on observations with the Mikhail Pavlinsky ART-XC telescope, hard X-ray instrument on board the SRG observatory. The SRG observatory was created by Roskosmos in the interests of the Russian Academy of Sciences represented by its Space Research Institute (IKI) in the framework of the Russian Federal Space Program, with the participation of Germany. The ART-XC team thanks the Roscosmos State Corporation, the Russian Academy of Sciences, and Rosatom State Corporation for supporting the ART-XC telescope, as well as the JSC Lavochkin Association and partners for manufacturing and running the Navigator spacecraft and platform.  
This work made also use of data supplied by the UK \textit{Swift} Science Data Centre at the University of Leicester and data obtained with NuSTAR mission, a project led by Caltech, funded by NASA, and managed by JPL. 
This research also has made use of the \nustar Data Analysis Software (NUSTARDAS) jointly developed by the ASI Science Data Centre (ASDC, Italy) and Caltech. We are grateful to the \swift and \nustar teams for approving and rapid scheduling of the monitoring campaign. This work made use of Astropy:\footnote{\url{http://www.astropy.org}} a community-developed core Python package and an ecosystem of tools and resources for astronomy \citep{astropy:2013, astropy:2018, astropy:2022}.
This research has made use of data and software provided by the High Energy Astrophysics Science Archive Research Centre (HEASARC), which is a service of the Astrophysics Science Division at NASA/GSFC and the High Energy Astrophysics Division of the Smithsonian Astrophysical Observatory. 

\end{acknowledgements}

\bibliography{allbib}
\bibliographystyle{aa}
\end{document}